\documentclass[11pt,a4paper]{article}
\usepackage{jcappub}
\usepackage{bm}
\usepackage{amsmath}
\title{
Inflationary Magnetogenesis in $R^{2}$-Inflation after Planck 2015
}
\author[a,b]{Anwar Saleh AlMuhammad}

\affiliation[a]{Department of Physics and Astronomy, The University of Texas at San Antonio (UTSA),\\One UTSA Circle, San Antonio, TX 78249, USA}
\affiliation[b]{Qatif Astronomy Society (QAS), Awjam, Qatif, KSA}

\emailAdd{anwar@physics.utexas.edu}

\abstract{
We study the primordial magnetic field generated by the simple model ${f^2}FF$ in Starobinsky, ${R^2}$-inflationary, model. The scale invariant PMF is achieved at relatively high power index of the coupling function, $\left| \alpha  \right| \approx 7.44$. This model does not suffer from the backreaction problem as long as, the rate of inflationary expansion, $H$, is in the order of or less than the upper bound reported by Planck ($ \le 3.6 \times {10^{ - 5}}{M_{{\rm{Pl}}}}$) in both de Sitter and power law expansion, which show similar results. We calculate the lower limit of the reheating parameter, ${R_{{\rm{rad}}}} > 6.888$ in ${R^2}$-inflation. Based on the upper limit obtained from CMB, we find that the upper limits of magnetic field and reheating energy density as, ${\left( {{\rho _{{B_{{\rm{end}}}}}}} \right)_{{\rm{CMB}}}} < 1.184 \times {10^{ - 20}}M_{{\rm{Pl}}}^4$ and ${\left( {{\rho _{{\rm{reh}}}}} \right)_{{\rm{CMB}}}} < 8.480 \times {10^{ - 22}}M_{{\rm{Pl}}}^4$. However, the limits derived from the inflationary model are  ${\left( {{\rho _{{B_{{\rm{end}}}}}}} \right)_{{{\rm{R}}^2} - {\rm{inflation}}}} < 4.6788 \times {10^{ - 29}}M_{{\rm{Pl}}}^4$ and ${\left( {{\rho _{{\rm{reh}}}}} \right)_{{{\rm{R}}^2} - {\rm{inflation}}}} < 3.344 \times {10^{ - 30}}M_{{\rm{Pl}}}^4$. All of foregoing results are well more than the lower limit derived from WMAP7 for both large and small field inflation. By using the Planck inflationary constraints, 2015 in the context of ${R^2}$-inflation, the upper limit of reheating temperature and energy density for all possible values of ,${w _{{\rm{reh}}}}$ are respectively constrained as, ${T_{{\rm{reh}}}} < 4.32 \times {10^{13}}{\rm{GeV}}$ and ${\rho _{{\rm{reh}}}} < 3.259 \times {10^{ - 18}}M_{{\rm{Pl}}}^4$ at ${n_{\rm{s}}} \approx 0.9674$. This value of spectral index is well consistent with Planck, 2015 results. Adopting ${T_{{\rm{reh}}}}$, enables us to constrain the reheating e-folds number, ${N_{{\rm{reh}}}}$ on the range $1 < {N_{{\rm{reh}}}} < 8.3$, for $- 1/3 < {w _{{\rm{reh}}}} < 1$. By using the scale invariant PMF generated by ${f^2}FF$, we find that the upper limit of present magnetic field, ${B_0} < 8.058 \times {10^{ - 9}}{\rm{G}}$. It is in the same order of PMF, reported by Planck, 2015.
}

\keywords{inflation, primordial magnetic fields}
\arxivnumber{}

\begin{document}
\maketitle
\flushbottom

\section{Introduction}
\label{sec:intro}

Inflationary cosmology solves the fundamental problems of the Big Bang model \cite{1}. It provides an explanation for the large scale structure of the Universe, which is linked to the quantum fluctuation of the field of inflation, $\phi$. The implications of inflation have been consistent with the observational aspects of the Universe in the large scale. 

On the other hand, large scale magnetic fields are being observed in all kinds of galaxies and cluster of galaxies at wide range of redshifts. Some evidences for the presence of the magnetic fields in voids were reported \cite{2}-\cite{8}. Such detection strongly enhances the primordial origin of the magnetic fields (PMF). The latest constraints on the value of PMF were presented recently by Planck 2015 \cite{9}. Hence, the seed of magnetic field cannot be generated by the galactic dynamo in the absence of charged particles \cite{10}-\cite{12}. The generation of PMF is still an open question in cosmology, see the reviews of this subject \cite{13}-\cite{20}.  However, one of the most interesting cosmological models is the simple (${f^2}FF$)model \cite{21}-\cite{25}. It has gained more interest because it is a stable model under perturbation. (See [26] and references therein.) Also, it can lead to a scale invariant spectrum of PMF \cite{23}-\cite{25}. A scale invariance property explains why PMF is detected nearly in all scales of the universe.

The main problems with this model are: the backreaction problem, where the scale of the energy of the associated electric field with PMF can exceed the scale of inflation itself \cite{24}-\cite{28}, and the strong coupling between electromagnetic fields and charged matter at the beginning of inflation \cite{27}-\cite{28}. There are some computational and phenomenological ways, which were 
proposed to avoid these problems \cite{22}-\cite{28}. For example, as the amplitude of magnetic field spectrum reach around the scale invariant value ($\sim n{\rm{G}}$) during inflation, the total electromagnetic energy density is much less than the energy of inflation itself \cite{22}. However, the avenues, proposed to overcome these two problems, need more investigations in order to form a robust model, which is self-consistent in both classical and quantum regimes. 

The first problem is classical and may spoil the inflation. However, the second problem is of quantum nature and may lead to a huge coupling between the electromagnetism and the charged particles \cite{27}. Thus, if we define this model as a classical consistent one, it cannot be ruled out only because of a quantum effect like strong coupling. Nevertheless, proving the consistency between the quantum limit and the classical limit of the model will make it more trustable. For this reason, solving the classical problem (backreaction) associated with this model makes it self-consistent model. That is what we try to achieve as a result of this paper.  

The Lagrangian of a scalar (inflaton) field $\phi $ coupled to the gauged electromagnetic vector field ${A_\mu }$ can be written \cite{23}-\cite{25} as,
\begin{equation}
\label{eq:1}
{\cal L} =  - \;\sqrt { - g} \left( {\frac{1}{2}\left( {{\partial _\mu }\phi } \right)\left( {{\partial ^\mu }\phi } \right) + V\left( \phi  \right) + \frac{1}{4}{g^{\alpha \beta }}{g^{\mu \nu }}{f^2}\left( {\phi ,{\rm{ }}t} \right){F_{\mu \alpha }}{F_{\nu \beta }}} \right)
\end{equation}
where, ${F_{\nu \beta }} = {\partial _\nu }{A_\beta } - {\partial _\beta }{A_\nu }$ is the electromagnetic field tensor and $g$ is the determinant of the spacetime metric ${g_{\mu \nu }}$. The first term in the Lagrangian is the standard kinetic part of the scalar field, and the second term,$V\left( \phi  \right)$, is the potential, which specifies the model of inflation.

The Lagrangian of a pure electromagnetic field is of the form, $ - \frac{1}{4}{F_{\mu \nu }}{F^{\mu \nu }}$. In Eq.(\ref{eq:1}), we couple it to the scalar field through the unspecified function $f(\phi ,{\rm{ }}t)$. The main reason behind this coupling is to break the conformal symmetry of electromagnetism and hence prevent the dilution of the seed of magnetic field as it is generated in the inflation era. On other word, electromagnetic fields would not feel inflation. At the end of inflation, the coupling function, $f(\phi ,{\rm{ }}t) \to 1$, to retrieve the conformal electromagnetism. The last condition is important to decide the form of coupling function and to study PMF in post-inflationary phases.       

The standard model of inflation based on a single scalar field, such as quadratic,$V\left( \phi \right) \sim {\phi^2}$, quartic, $V\left( \phi  \right)\sim{\phi^4} $[29], ${R^2}$-inflation, $V\left( \phi  \right)\sim M^4\left(1-\exp{[-\sqrt{2/3} \phi/M_{PL}]}\right)^{2} $ [30, 41] and the exponential potential, $V\left( \phi  \right)\sim\exp [ - \sqrt {2{\epsilon _1}} \left( {\phi  - {\phi _0}} \right)]$[31]. The last one is used in \cite{23}-\cite{24} to find the magnetic and electric spectrum in the ${f^2}FF$ model. These models became more interesting after WMAP \cite{32}-\cite{33} and Planck \cite{34}-\cite{35}. 

Very recently, the joint analysis of BICEP2/Keck Array and Planck (BKP) data was released on Feb 2015. The joint data of three probes eliminate the effect of dust contamination and show that the upper limit of tensor to scalar ration, ${r_{0.05}} < 0.12$ at 95\% CL, and the gravitational lensing B-modes (not the primordial tensor) are detected in $7\sigma $ \cite{36}. Similarly, the inflationary models are constrained based on the new analysis of data. The scalar spectral index was constrained by Planck, 2015 to be ${n_s} = 0.9682 \pm 0.0062$ \cite{37}. As a result, the more standard inflationary models, like ${R^2}$-inflation, which result low value of $r$, are the most favored one. However, the chaotic inflationary models like large field inflation (LFI) and natural inflation (NI) are disfavored \cite{37}. These results ruled out the first results of BICEP2, 2014 \cite{38}. 

In this paper, the simple model, ${f^2}FF$, of PMF will be investigated in the context of ${R^2}$-inflation, in the same way we did for NI and LFI, \cite{39}-\cite{40}. Further, we will constrain the reheating parameters under the same inflationary model by using the reported upper limit of PMF by Planck, 2015. The present PMF will be constrained based on the scale invariant magnetic field generated during inflationary era.

Throughout this paper, we adopt the natural units, $\left[ {c = \hbar  = {k_B} = \;1} \right]$, the signature $( - 1,{\rm{ 1, 1, 1)}}$, and flat universe, where we use the reduced Planck mass, ${M_{{\rm{Pl}}}}{\rm{\;}} = {\left( {8\pi G} \right)^{ - 1/2}}$. It will be taken, ${M_{{\rm{Pl}}}}{\rm{\;}} = 1$ in the computation parts. Hence, the potential of ${R^2}$-inflation (Starobinsky model) can be written in Einstein frame \cite{41}-\cite{42} as, 
\begin{equation}
\label{eq:2}
V\left( \phi  \right)= M^4\left(1-\exp{[-\sqrt{2/3} \phi/M_{PL}]}\right)^{2} 
\end{equation} 
where, $M$ is the amplitude of the potential and it can be determined by the amplitude of CMB anisotropies. 

The order of this paper will be as follows, in section.\ref{sec:slow roll}, the slow roll inflation formulation is presented for both simple de Sitter model of expansion and the more general power law expansion in the context of ${R^2}$-inflation. In section.\ref{sec:EM in R2}, the PMF and associated electric fields are computed.  The constraining of reheating parameters in the same model of inflation and by using the reported results of Planck, 2015 are discussed in section.\ref{sec:Reh by PMF}. In section.\ref{sec:PMF from Magnetogens}, the present PMF is constrained based on the computed magnetogensis. In section.\ref{sec:summary}, a summary of the results is presented.

\section{Slow roll analysis of ${R^2}$-inflation}
\label{sec:slow roll}

During inflation, we will assume the electromagnetic field to be negligible compared to the scalar field, $\phi $ \cite{24}. Hence, the equation of motion derived from \ref{eq:1} for the scalar field can be written as,
\begin{equation}
\label{eq:3}
\ddot \phi  + 3H\dot \phi  + {V_\phi } = 0 
\end{equation} 
where, $H\left( t \right) = \;\dot a\left( t \right)/a\left( t \right)$, is the Hubble parameter as a function of cosmic time, $t$, and $a\left( t \right)$ is the cosmological scale factor. The over dot indicates differentiation with respect to cosmic time, and ${V_\phi } = {\partial _\phi }V$. The Friedman equation can be obtained from the Einstein field equations by assuming a spatially-flat Friedmann-Robertson-Walker FRW universe, which yields,
\begin{equation}
\label{eq:4}
H{\;^2} = \;\frac{1}{{3{M_{{\rm{Pl}}}}^2}}\left[ {\frac{1}{2}{{\dot \phi }^2} + V\left( \phi  \right)} \right] - \frac{K}{{{a^2}}} + \frac{{\rm{\Lambda }}}{3}
\end{equation}
where, $K$, is the curvature of the universe and, ${\rm{\Lambda }}$, is the cosmological constant. The last two terms of \ref{eq:4} can be neglected in the inflation era. Also, under slow roll approximation, one can neglect the second derivative in (\ref{eq:3}), which leads to the attractor condition,
\begin{equation}
\label{eq:5}
\dot \phi  \simeq  - \frac{{{V_\phi }}}{{3H}}
\end{equation}
 
Defining the slow roll parameters of inflation in terms of the potential \cite{30}-\cite{43}, of the ${R^2}$-inflation,
\begin{equation}
\label{eq:6}
{\epsilon _{1V}}\left( \phi  \right) = \frac{1}{2}{M_{{\rm{Pl}}}}^2{\left( {\frac{{{V_\phi }}}{V}} \right)^2} = \frac{4}{3}{\left[ { - 1 + {\rm{exp}}\left( {\sqrt {\frac{2}{3}} \frac{\phi }{{{M_{{\rm{Pl}}}}}}} \right)} \right]^2},
\end{equation}

\begin{equation}
\label{eq:7}
{\epsilon _{2V}}\left( \phi  \right) = {M_{{\rm{Pl}}}}^2\left[ {{{\left( {\frac{{{V_\phi }}}{V}} \right)}^2} - \frac{{{V_{\phi \phi }}}}{V}} \right] = \frac{2}{3}{\left[ {{\rm{sinh}}\left( {\frac{\phi }{{\sqrt 6 {M_{{\rm{Pl}}}}}}} \right)} \right]^{ - 2}},
\end{equation}
 
\begin{equation}
\label{eq:8}
{\epsilon _{3V}}\left( \phi  \right) = \frac{2}{3}\left( {\coth \left[ {\frac{\phi }{{\sqrt 6 {M_{{\rm{Pl}}}}}}} \right] - 1} \right)\coth \left[ {\frac{\phi }{{\sqrt 6 {M_{{\rm{Pl}}}}}}} \right].
\end{equation}
They also, can be written in terms of the Hubble parameter,
\begin{equation}
\label{eq:9}
{\epsilon _{1H}}\left( \phi  \right) = 2{M_{{\rm{Pl}}}}^2{\left( {\frac{{{H_\phi }}}{H}} \right)^2},{\rm{    }}{\epsilon _{2H}}\left( \phi  \right) = 2{M_{{\rm{Pl}}}}^2\left( {\frac{{{H_{\phi \phi }}}}{H}} \right).
\end{equation}

Finally, the relation between the two formalisms \cite{43}-\cite{44} can be written as
\begin{equation}
\label{eq:10}
{\epsilon _{1V}} = {\epsilon _{1H}}{\left( {\frac{{3 - {\epsilon _{2H}}}}{{3 - {\epsilon _{1H}}}}} \right)^2}.
\end{equation}
All of the above parameters are assumed to be very small during the slow roll inflation,$({\epsilon _{1V}}$, ${\epsilon _{2V}}$,${\epsilon _{1H}}$, and ${\epsilon _{2H}}) \ll 1$. Further, inflation ends when, the values of $({\epsilon _{1V}}{\rm{ }},{\rm{ }}{\epsilon _{1H}}) \to 1$. In the first order of approximation, one can neglect ${\epsilon _{1H}}$ and ${\epsilon _{2H}}$comparing with 3, obtaining ${\epsilon _{1V}} \simeq {\epsilon _{1H}}$. Therefore, using (\ref{eq:9}) and the relation between the cosmic time, $t$, and the conformal time,$\eta $, $dt = a\left( \eta  \right)d\eta $, one can write the relation between conformal time and slow roll parameter, ${\epsilon _{1H}}$, as \cite{45},

\begin{equation}
\label{eq:11}
\eta  =  - \frac{1}{{aH}} + \int {\frac{{{\epsilon _{1H}}}}{{{a^2}H}}da}.
\end{equation}
Assuming, ${\epsilon _{1H}} \approx const$, and then integrating (\ref{eq:11}) yields the power law expansion of the universe during inflation,
\begin{equation}
\label{eq:12}
a(\eta ) = l{\left| \eta  \right|^{ - 1 - {\epsilon _{1H}}}} \simeq l{\left| \eta  \right|^{ - 1 - {\epsilon _{1V}}}},
\end{equation}
where, $l$ is integration constant. 

On the other hand, the universe expands exponentially during inflation at a very high but almost constant rate, $H$, in the simplest form of inflationary expansion (de Sitter),

\begin{equation}
\label{eq:13}
H = \frac{{\dot a}}{a} \simeq {\rm{const}},
\end{equation}
  
\begin{equation}
\label{eq:14}
a\left( t \right) = a\left( {{t_1}} \right)\;{\rm{exp}}\left[ {{H_i}t} \right],
\end{equation}
where, ${t_1}$, is the start time of inflation. In a conformal time, Eq.(\ref{eq:14}) can be written as,

\begin{equation}
\label{eq:15}
a\left( \eta  \right) =  - \frac{1}{{H\;\eta \; + \;{c_1}}},
\end{equation}
where, ${c_1}$, is the integration constant. Plugging (\cite{15}) into the relation between cosmic and conformal time and integrating, implies that $\eta  \to \left( { - \infty ,{\rm{ }}{0^ - }} \right)$ as $t \to \left( {0,{\rm{ }}\infty } \right)$. Thus, if $\eta  \to {0^ - }$, then $a\left( \eta  \right) \to \infty $ and ${c_1} \to 0$. Therefore, we can write $a\left( \eta  \right) =  - 1/\left( {H\;\eta } \right)$.

\begin{figure}[tbp]
\centering  \begin{center}

\includegraphics[width=0.7\textwidth,origin=c]{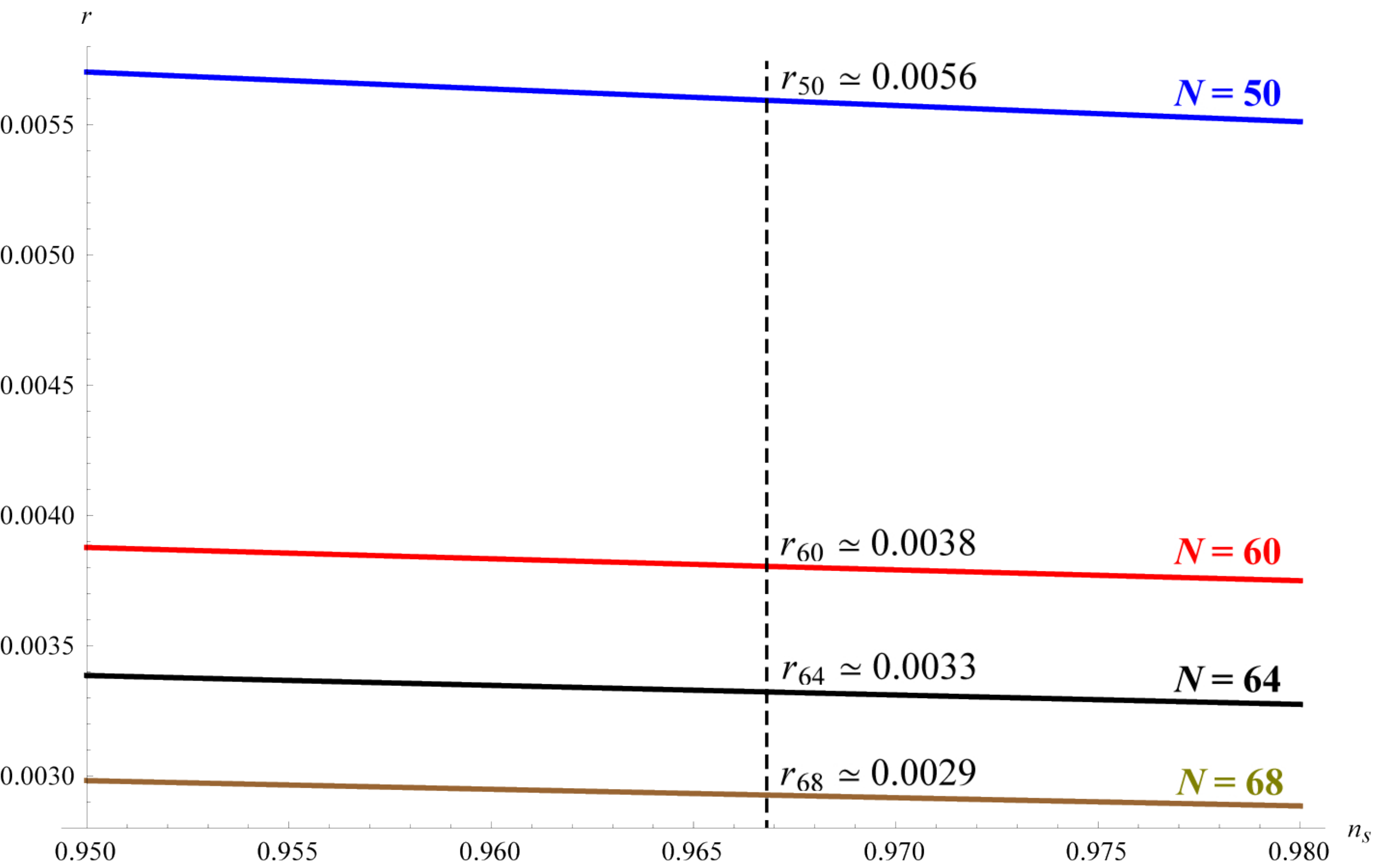}

\caption{\label{fig:1} The $r-n_{s}$ relation in $R^2$-inflation for $N = 50,60,64$, and 68. All values of ${r_N}$ at ${n_s} \simeq 0.968$, are well below upper limit of $r (< 0.12)$ by BKP, 2015.}
\end{center}
\end{figure}

Also, the relation between slow roll parameters and the scalar power spectrum amplitude,${A_s}$, the tensor power spectrum amplitude,${A_t}$, the scalar spectral index,${n_s}$, and tensor-to-scalar ratio, $r$, can be written as follows \cite{43}-\cite{46},

\begin{equation}
\label{eq:16}
{A_s} = \frac{V}{{24{\pi ^2}{M_{{\rm{Pl}}}}^4{\epsilon _{1V}}}},
\end{equation}

\begin{equation}
\label{eq:17}
{A_t} = \frac{{2V}}{{3{\pi ^2}{M_{{\rm{Pl}}}}^4}},
\end{equation} 
	 
\begin{equation}
\label{eq:18}
{n_s} = 1 - 6{\epsilon _{1V}} + 2{\epsilon _{2V}} = 1 - 8{\left( {{{\rm{e}}^{\sqrt {\frac{2}{3}} \frac{\phi }{{{M_{{\rm{Pl}}}}}}}} - 1} \right)^{ - 2}} + \frac{4}{3}{\rm{csch}}{\left[ {\frac{\phi }{{\sqrt 6 {M_{{\rm{Pl}}}}}}} \right]^2} \approx 1 + \frac{4}{3}{\rm{csch}}{\left[ {\frac{\phi }{{\sqrt 6 {M_{{\rm{Pl}}}}}}} \right]^2},
\end{equation}

\begin{equation}
\label{eq:19}
r = \frac{{{A_t}}}{{{A_s}}} = 16{\epsilon _{1V}} = \frac{{64}}{3}{\left( {{{\rm{e}}^{\sqrt {\frac{2}{3}} \frac{\phi }{{{M_{{\rm{Pl}}}}}}}} - 1} \right)^{ - 2}}.
\end{equation}
One can find the relation between $r$ and ${n_s}$ which depends on the number of e-folds of inflation, $N$. The first order of approximation for $N$ can be written as,

\begin{equation}
\label{eq:20}
N = \ln \left( {\frac{{a(\eta )}}{{a({\eta _i})}}} \right) \simeq  - \sqrt {\frac{1}{{2{M_{{\rm{Pl}}}}^2}}} \;\mathop \smallint \limits_\phi ^{{\phi _{{\rm{end}}}}} \frac{1}{{\sqrt {{\epsilon _1}} }}d\phi \; \simeq \frac{3}{4}\exp \left( {\sqrt {\frac{2}{3}} \frac{\phi }{{{M_{{\rm{Pl}}}}}}} \right),
\end{equation}
where, $\phi $, is the initial field and ${\phi _{\rm{end}}}$ is the field at the end of inflation. Solving for $\phi $ from (\ref{eq:20}),

\begin{equation}
\label{eq:21}
\phi  \simeq \sqrt {\frac{3}{2}} {M_{{\rm{Pl}}}}\ln \left( {\frac{{4({N_{{\rm{end}}}} - N)}}{3}} \right).
\end{equation}
where, ${N_{{\rm{end}}}}$ is the number of e-folds at the end of inflation. For simplicity, we will denote $({N_{{\rm{end}}}} - N) \equiv N$. Hence, plugging (\ref{eq:21}) into (\ref{eq:18})-(\ref{eq:19}), yields,
\begin{equation}
\label{eq:22}
r \simeq \frac{{192}}{{{{\left( {3 - 4N} \right)}^2}{n_s} - 64N}}.
\end{equation}
The relation (\ref{eq:22}) is drawn for some interesting values of $N$ in Fig.\ref{fig:1}. The values of $r$, are well below the limit of BKP, 2015 ($r < 0.12$).

\section{The electromagnetic spectra generated in $R^{2}$-inflation}
\label{sec:EM in R2}

One can write the electromagnetic equation, for ${A_\mu }$from the Lagrangian, (\ref{eq:1}), as,
\begin{equation}
\label{eq:23}
{\partial _\mu }\left[ {\sqrt { - g} {g^{\mu \nu }}{g^{\alpha \beta }}{f^2}\left( {\phi ,t} \right){F_{\nu \beta }}} \right] = 0,
\end{equation}
where, $g$ is the determinant of the flat FRW space-time metric ${g_{\mu \nu }}$ defined as,
\begin{equation}
\label{eq:24}
d{s^2} = {g_{\mu \nu }}d{x^\mu }\;d{x^\nu } =  - d{t^2} + {a^2}\left( t \right)d{x^2} = {g'_{\mu \nu }}d{x^\mu }\;d{x^\nu } =  - {a^2}\left( \eta  \right)\left( {d{\eta ^2} + d{x^2}} \right).
\end{equation}
Hence, $\sqrt { - g}  = {a^3}\left( t \right)$, $\sqrt { - g'}  = {a^4}\left( \eta  \right)$, and ${g^{\mu \nu }}{g_{\mu \beta }} = \delta _\beta ^\nu $. Adopting the Coulomb (radiation) gauge, ${\partial _i}{A^i}\left( {t,x} \right) = 0$, and the charge-free condition ${A_0}\left( {t,x} \right) = 0$, Eq.(\ref{eq:23}) can be written as
\begin{equation}
\label{eq:25}
{\ddot A_i}\left( {t,x} \right) + \left( {H + \frac{{2\dot f}}{f}} \right){\dot A_i}\left( {t,x} \right) - {\rm{\;}}{\partial _j}{\partial ^j}{A_i}\left( {t,x} \right) = 0.
\end{equation}
In conformal time, the above equation can be written as 
\begin{equation}
\label{eq:26}
{{\rm{A''}}_i}\left( {\eta ,x} \right) + 2\frac{{f'}}{f}{{\rm{A'}}_i}\left( {\eta ,x} \right) - {a^2}\left( \eta  \right){\rm{\;}}{\partial _j}{\partial ^j}{{\rm{A}}_i}\left( {\eta ,x} \right) = 0.
\end{equation}
Define the function, ${{\rm{\bar A}}_i}\left( {\eta ,x} \right) = f\left( \eta  \right){{\rm{A}}_i}\left( {\eta ,x} \right)$, and plug it into (\ref{eq:26}),
\begin{equation}
\label{eq:27}
\overline {{{{\rm{A''}}}_i}}  + \frac{{f''}}{f}{{\rm{\bar A}}_i} - \;{a^2}{\rm{\;}}{\partial _j}{\partial ^j}{{\rm{\bar A}}_i}\left( {\eta ,x} \right) = 0.
\end{equation} 
	
On the other hand, the quantization of ${{\rm{\bar A}}_i}$ can be written in terms of creation and annihilation operators, ${b^\dag }_\lambda $ and ${b_\lambda }\left( k \right)$, as,
\begin{equation}
\label{eq:28}
{{\rm{\bar A}}_i}\left( {\eta ,x} \right) = \int {\frac{{{d^3}k}}{{{{\left( {2\pi } \right)}^{3/2}}}}\mathop \sum \limits_{\lambda  = 1}^2 {\epsilon _{i\lambda }}\left( k \right)[{b_\lambda }\left( k \right){\cal A}\left( {\eta ,k} \right){e^{ik.x}} + {b^\dag }_\lambda \left( k \right){{\cal A}^*}\left( {\eta ,k} \right){e^{ - ik.x}}]},
\end{equation}
where, ${\epsilon _{i\lambda }}$ is the transverse polarization vector, and $k = \frac{{2\pi }}{\lambda }$, is the commoving wave number. Substituting (\ref{eq:28}) into (\ref{eq:27}),
\begin{equation}
\label{eq:29}
{\cal A}''\left( {\eta ,k} \right) + \left( {{k^2} - \frac{{f''}}{f}} \right){\cal A}\left( {\eta ,k} \right) = 0,
\end{equation}
where we have used, ${\partial ^j} = {g^{jk}}{\partial _k} = \left( {\frac{{{\delta ^{jk}}}}{{{a^2}}}} \right){\partial _k}$. 
Similarly, the stress energy tensor, ${T_{\mu \nu }}$, can be written in terms of the action, $S$,  as,
\begin{equation}
\label{eq:30}
{T_{\mu \nu }} =  - \frac{2}{{\sqrt { - g} }}\frac{{\delta S}}{{\delta {g^{\mu \nu }}}} = \left( {{\partial _\mu }\phi } \right)\left( {{\partial _\nu }\phi } \right) - {f^2}\left( {\phi ,{\rm{ }}t} \right){g^{\alpha \beta }}{F_{\mu \alpha }}{F_{\beta \nu }} - \frac{1}{4}{g_{\mu \nu }}{g^{\alpha \beta }}{g^{\gamma \delta }}{f^2}\left( {\phi ,{\rm{ }}t} \right){F_{\beta \delta }}{F_{\alpha \gamma }}.
\end{equation}
For the magnetic field, we ignore the kinetic part (first term) because it does not contribute to the electromagnetic field. Then magnetic part of ${T_{\mu \nu }}$,
\begin{equation}
\label{eq:31}
T_{00}^B = \frac{1}{4}{a^2}{g^{ij}}{g^{kl}}{f^2}\left( {\phi ,{\rm{ }}t} \right)({\partial _j}{A_l} - {\partial _l}{A_j})({\partial _i}{A_k} - {\partial _k}{A_i}) = \frac{{{f^2}\left( {\phi ,{\rm{ }}t} \right)}}{2}{a^2}{B_\mu }{B^\mu }.
\end{equation}

The energy density of the magnetic field, ${\rho _B}$ is found by taking the average of the component of the stress energy tensor. Finally, the spectrum of the magnetic field is given \cite{24} by 
\begin{equation}
\label{eq:32}
\frac{{d{\rho _B}}}{{d{\rm{ln}}k}} = \frac{1}{{2{\pi ^2}}}{\left( {\frac{k}{a}} \right)^4}k{\left| {{\cal A}\left( {\eta ,k} \right)} \right|^2}.
\end{equation}
Similarly, the spectrum of the electric field is given by
\begin{equation}
\label{eq:33}
\frac{d{\rho_E}}{d{\rm{ln}}k} = \frac{f^2}{2\pi ^2}\frac{{{k^3}}}{{{a^4}}}{\left| {{\left[ \frac{{\cal A}\left( {\eta ,{\rm{ }}k} \right)}{f} \right]}'} \right|^2}.
\end{equation}
Therefore, to calculate the electromagnetic spectra, one has to find ${\cal A}\left( {\eta ,{\rm{ }}k} \right)$, which can be achieved by solving (\ref{eq:29}). But first, we need to specify the coupling function, $f\left( \phi  \right)$. 
Assuming that the relation between the coupling function and scale factor is of the power law form \cite{8}, $f\left( \eta  \right){\rm{\;}} \propto {\rm{\;}}{a^\alpha }$, where, $\alpha $ is free index to be determined later from the shape of the spectrum of PMF. Then, by combining (\ref{eq:3}) and (\ref{eq:4}) in the slow roll limit,
\begin{equation}
\label{eq:34}
f\left( \phi  \right){\rm{\;}} \propto exp\left[ { - \frac{\alpha }{{3\;{M_{{\rm{Pl}}}}^2}}\mathop \smallint \limits_{}^\phi  \frac{V}{{{V_\phi }}}d\phi } \right].
\end{equation}
Substituting (\ref{eq:21}) into (\ref{eq:2}) and (\ref{eq:34}) yields the coupling function as a function of $N$,
\begin{equation}
\label{eq:35}
f\left( N \right){\rm{\;}} = D\left[ {{{\left( {\frac{{4N}}{3}} \right)}^{\frac{\alpha }{4}}}{{\rm{e}}^{ - \frac{{N\alpha }}{3}}}} \right],
\end{equation}
where, $D$ is a coupling constant. Substituting (\ref{eq:35}) into (\ref{eq:29}) gives,
\begin{equation}
\label{eq:36}
{\cal A}''\left( {\eta ,k} \right) + \left( {{k^2} - Y\left( \eta  \right)} \right){\cal A}\left( {\eta ,k} \right) = 0,
\end{equation} 
here the function $Y\left( \eta  \right) = \frac{{f''}}{f}$. Hence, we need to write the derivative in terms of e-folds number, $N$. In the next two sections, the electromagnetic spectra are calculated for both de Sitter and power law expansion.

\subsection{Inflationary electromagnetic spectra in de Sitter expansion}
\label{subsec:EM in dS}   

Shortly after the onset of inflation the value of $H$ becomes very high and is approximately constant, but later on, it decreases as the value of the field changes. For the zeroth approximation and after the first few e-foldings, we can consider H as a constant ratio. That is basically the de Sitter expansion, which is exactly exponential expansion as described by (\ref{eq:14}). In fact, de Sitter model does not have graceful exit from inflation \cite{46}. But it can be assumed as a valid approximation on part of the inflation. 
Assuming, $a({\eta _i}) = {\rm{const}}$, and substituting of (\ref{eq:15}) into (\ref{eq:20}) and differentiating both sides, yields that, ${\partial _{\eta \eta }}f(\eta ) = {\eta ^{ - 2}}{\partial _{NN}}f(N)$. Therefore, the function $Y\left( \eta  \right)$ can be written as, 
\begin{equation}
\label{eq:37}
{Y_{dS}}\left( \eta  \right) = \frac{{\alpha \left[ {{{\left( {3 - 4N} \right)}^2}\alpha  - 36} \right]}}{{144{N^2}{\eta ^2}}}.
\end{equation}
Also, as the variation in $N$ is very small comparison with the variation in $\eta $, we will assume that $N$ is quasi-constant and we will not write it as a function of conformal time explicitly in Eq.(36). Therefore, substituting of (\ref{eq:37}) into (\ref{eq:36}), yields,
\begin{equation}
\label{eq:38}
{\cal A}\left( {\eta ,k} \right) = {\left( {k\eta } \right)^{1/2}}\left[ {{C_1}\left( k \right)\;{J_\chi }\left( {k\eta } \right) + {C_2}\left( k \right)\;{J_{ - \chi }}\left( {k\eta } \right)} \right],
\end{equation}
where, ${J_\chi }\left( {k\eta } \right)$, is the Bessel function of the first kind, and the argument of the function, 
\begin{equation}
\label{eq:39}
{\chi _{dS}}(\alpha ,N) = \frac{{\sqrt {36{N^2} - 36\alpha  + {{\left( {3 - 4N} \right)}^2}{\alpha ^2}} }}{{12N}}.
\end{equation}

\begin{figure}[tbp]
\centering  \begin{center}

\includegraphics[width=0.6\textwidth,origin=c]{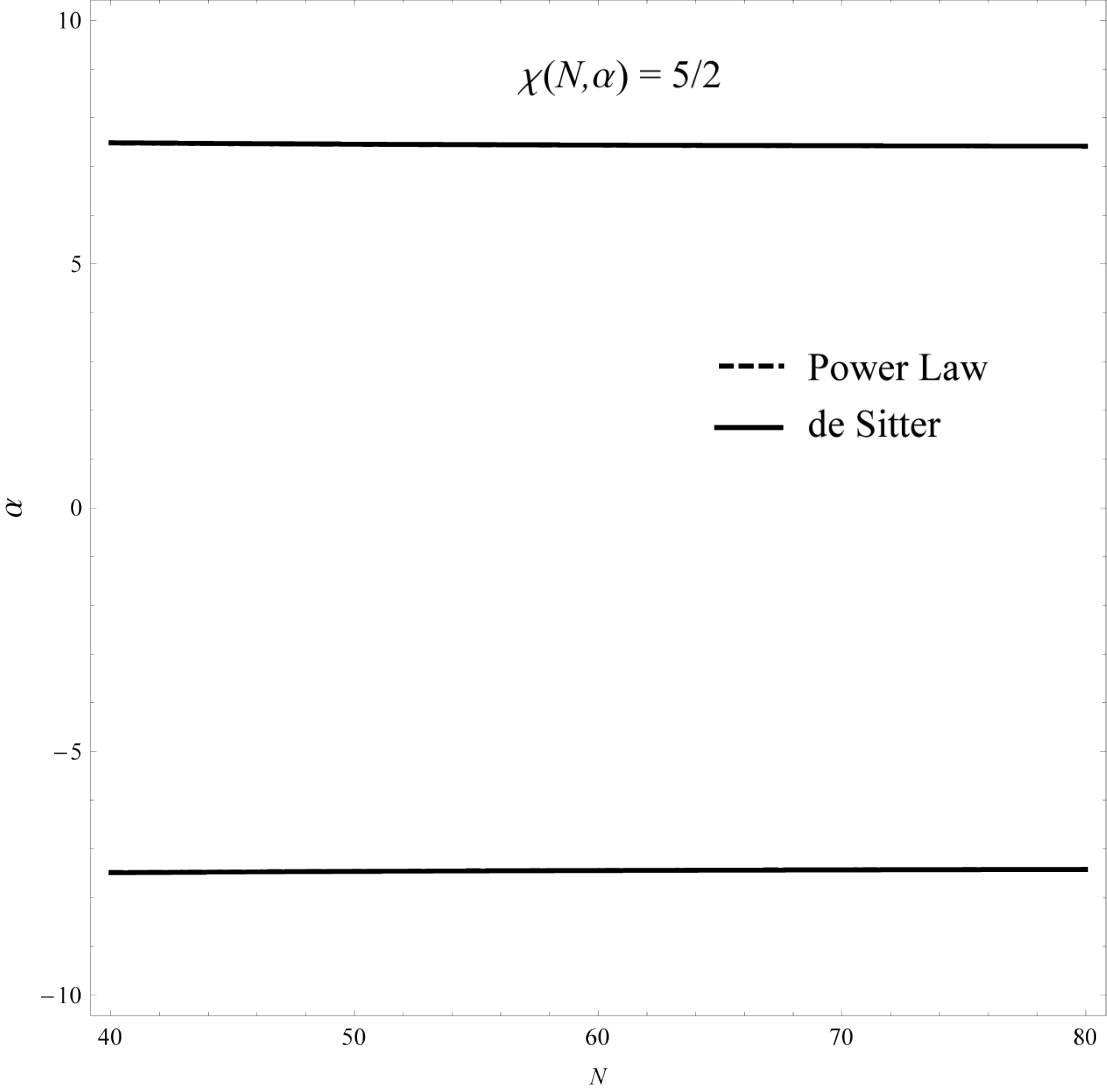}

\caption{\label{fig:2} The $\alpha  - N$ relation, at $\chi (N,\alpha ) = 5/2$, the case at which we can generate the scale invariant PMF.}
\end{center}
\end{figure}

The relevant PMF is obtained in the long wavelength regime, $k \ll 1$ (outside Hubble radius). In this limit, Eq.(\ref{eq:38}) can be written as \cite{24},
\begin{equation}
\label{eq:40}
{{\cal A}_{k \ll 1}}\left( {\eta ,k} \right) = {\left( k \right)^{ - 1/2}}\left[ {{D_1}\left( \chi  \right)\;{{\left( {k\eta } \right)}^{\chi  + \frac{1}{2}}} + {D_2}\left( \chi  \right){{\left( {k\eta } \right)}^{\frac{1}{2} - \chi }}} \right].
\end{equation}
The constants, $\;{D_1}\left( \chi  \right)$ and ${D_2}\left( \chi  \right)$, can be fixed by using the normalization of ${\cal A}\left( {\eta ,{\rm{ }}k} \right)$ and the other limit, \[{{\cal A}_{k \gg 1}}\left( {\eta ,{\rm{ }}k} \right) \to {{{e^{ - k\eta }}} \mathord{\left/
 {\vphantom {{{e^{ - k\eta }}} {\sqrt {2k} }}} \right.
 \kern-\nulldelimiterspace} {\sqrt {2k} }}\]. Thus, they can be written as
\begin{equation}
\label{eq:41}
{D_1}\left( \chi  \right) = \frac{{\sqrt \pi  }}{{{2^\chi }}}\;\frac{{{e^{ - i\pi (\chi  + \frac{1}{2})/2}}}}{{{\rm{\Gamma }}\left( {\chi  + 1} \right)\cos \left( {\pi (\chi  + \frac{1}{2})} \right)}},{\rm{   }}{D_2}\left( \chi  \right) = \frac{{\sqrt \pi  }}{{{2^{1 - \chi }}}}\;\frac{{{e^{ - i\pi \left( {\chi  + \frac{3}{2}} \right)/2}}}}{{{\rm{\Gamma }}\left( {1 - \chi } \right)\cos \left( {\pi (\chi  + \frac{1}{2})} \right)}}.
\end{equation}
By substituting (\ref{eq:40}) into (\ref{eq:32})-(\ref{eq:33}), we get the spectra of the magnetic and electric fields. If the PMF is scale invariant, then the magnetic spectrum should be constant, which could be achieved at $\chi  = 5/2$. This value can only be obtained around, $\alpha  \approx \{  - 7.44,\;{\rm{ }}7.44\} $, see Fig.\ref{fig:2}.

\begin{figure}[tbp]
\centering  \begin{center}
\includegraphics[width=0.8\textwidth,origin=c]{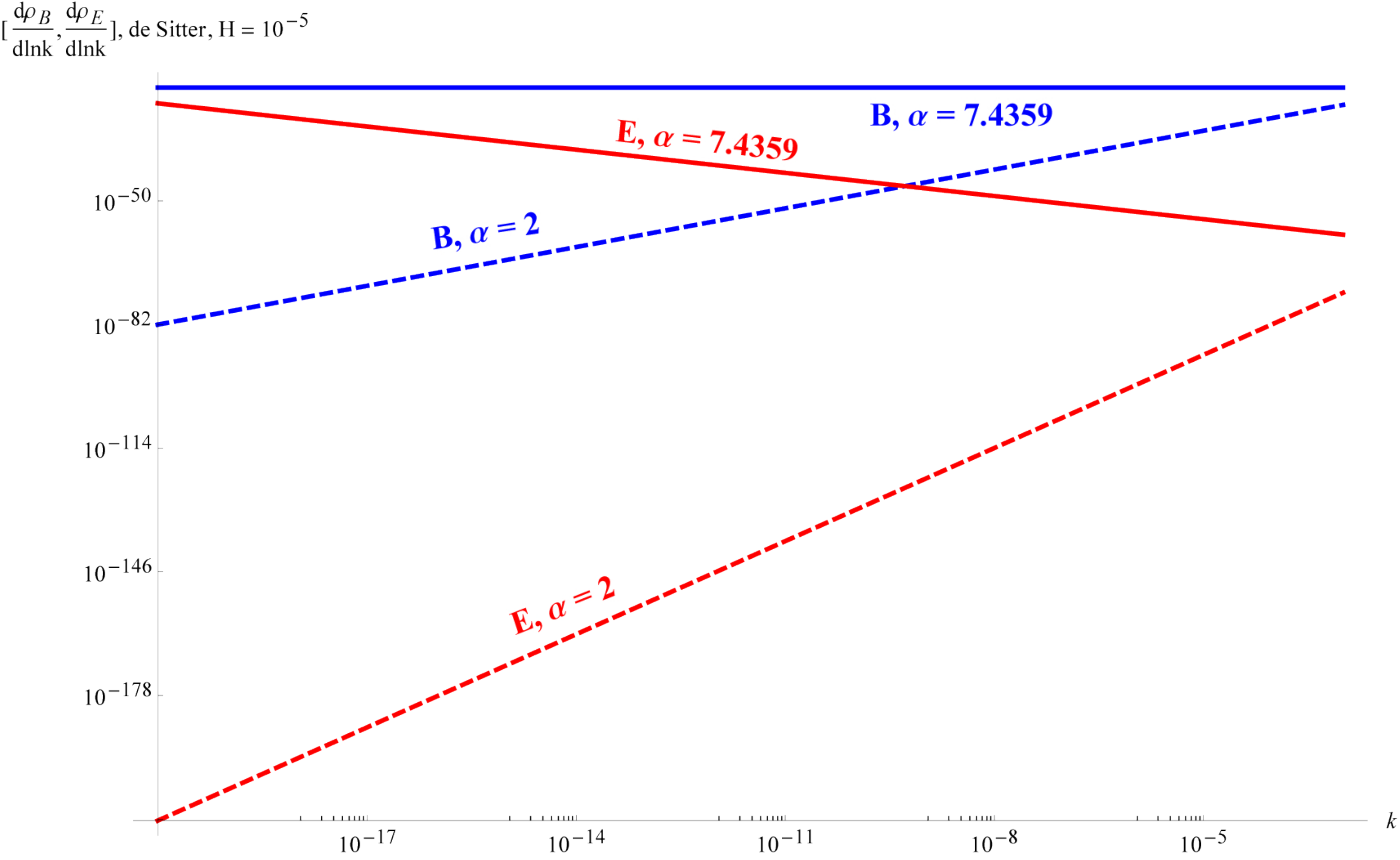}

\caption{\label{fig:3} The EM spectra for  $\alpha  = \left[ 2,7.4359 \right]$, $N = 64$,  $\eta  =  - {10^{ - 2}}$ and the expansion rate , $H = {10^{ - 5}}{M_{{\rm{Pl}}}}$. For $\alpha  = 2$, the scale invariant PMF cannot be achieved although the backreaction problem is easily to be avoided. However, for $\chi (N,\alpha ) = 5/2$ ($\alpha  = 7.4359$), the scale invariant PMF can be generated and the magnetic field energy is less than the electric field which can avoid the backreaction problem.}
\end{center}
\end{figure}

Although these values of $\alpha$ are too high, which may exceed dynamo limit \cite{24}, we adopt the positive value, $\alpha  \approx 7.44$. In the last section, we will discuss this point. So, let us assume it is valid at this point and use it to investigate the electromagnetic spectra generated by ${f^2}FF$ for long wavelength approximation ($k \ll 1$). From (\ref{eq:32})-(\ref{eq:33}), one can plot the electromagnetic spectra for different values of $H$ and $\alpha $, see Fig.\ref{fig:3}, for $H = {10^{ - 5}}{M_{{\rm{Pl}}}}$, which is around the Planck, 2015 upper limit of pivot Hubble parameter, (${H_*} < 3.6 \times {10^{ - 5}}{M_{{\rm{Pl}}}}$) \cite{37}. The pivot moment is the time when the commoving scale (${k_*} = {a_*}{\rm{ }}{H_*} \simeq 0.05Mp{c^{ - 1}}$) exits the Hubble radius. In this case, the scale invariant PMF can be generated and the backreaction problem may be avoided at $\alpha  \simeq 7.4359$. However, for $\alpha  = 2$, the backreaction problem can be avoided easily, but the scale invariant PMF is not generated, see Fig.\ref{fig:3}.

\begin{figure}[tbp]
\centering  \begin{center}
\includegraphics[width=0.8\textwidth,origin=c]{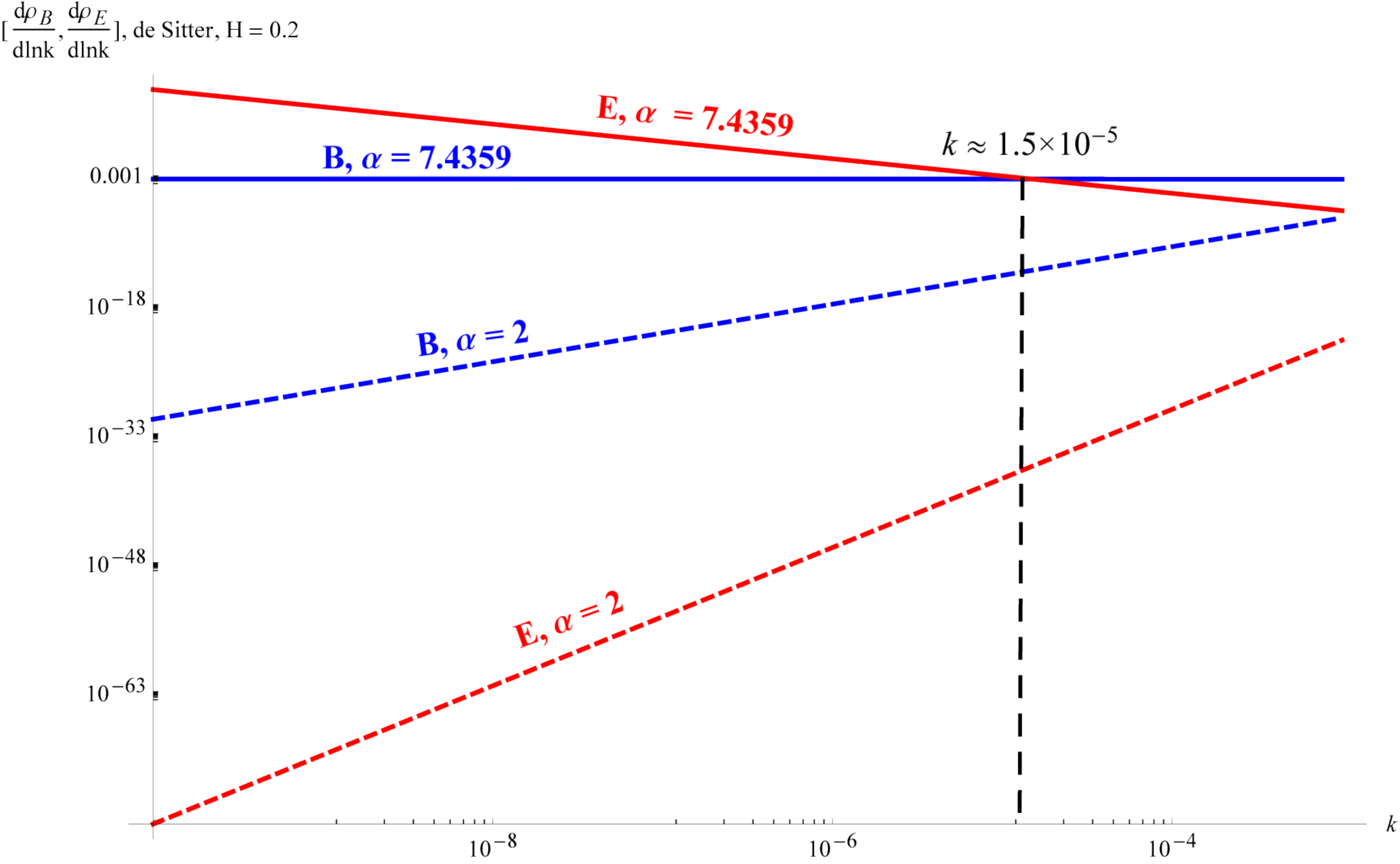}

\caption{\label{fig:4} The EM spectra for $\alpha  = \left[ 2, 7.4359 \right]$, $N = 64$, $\eta  =  - {10^{ - 2}}$ and the expansion rate,  $H = 0.2{M_{{\rm{Pl}}}}$. For $\alpha  = 2$, the scale invariant PMF cannot be achieved although the backreaction problem is easily to be avoided. However, for $\chi (N,\alpha ) = 5/2$ ($\alpha  \simeq 7.4359$), the scale invariant PMF can be generated but the magnetic field energy is less than the electric field for low wavenumber. The PMF exceeds the electric fields at, $k \sim 1.5 \times {10^{ - 5}}$.}
\end{center}
\end{figure}

On the other hand, in the large expansion rate case, $H = 0.2{M_{{\rm{Pl}}}}$, the scale invariant PMF can be generated at $\alpha  \simeq 7.4359$ but the problem of backreaction may not be avoided in the low, $k \ll 1$, see Fig.\ref{fig:4}. In fact, this value of $H$ is way above the constraint upper limit \cite{37}, but we use it to decide the scale at which the electric field exceeds magnetic field ($k \sim 1.5 \times {10^{ - 5}}$). Hence, we can use this value of the wavenumber as an upper limit in this case. Again, for $\alpha  = 2$, the backreaction problem can be avoided easily, but the scale invariant PMF is not maintained throughout the inflationary era.

\begin{figure}[tbp]
\centering  \begin{center}
\includegraphics[width=0.6\textwidth,origin=c]{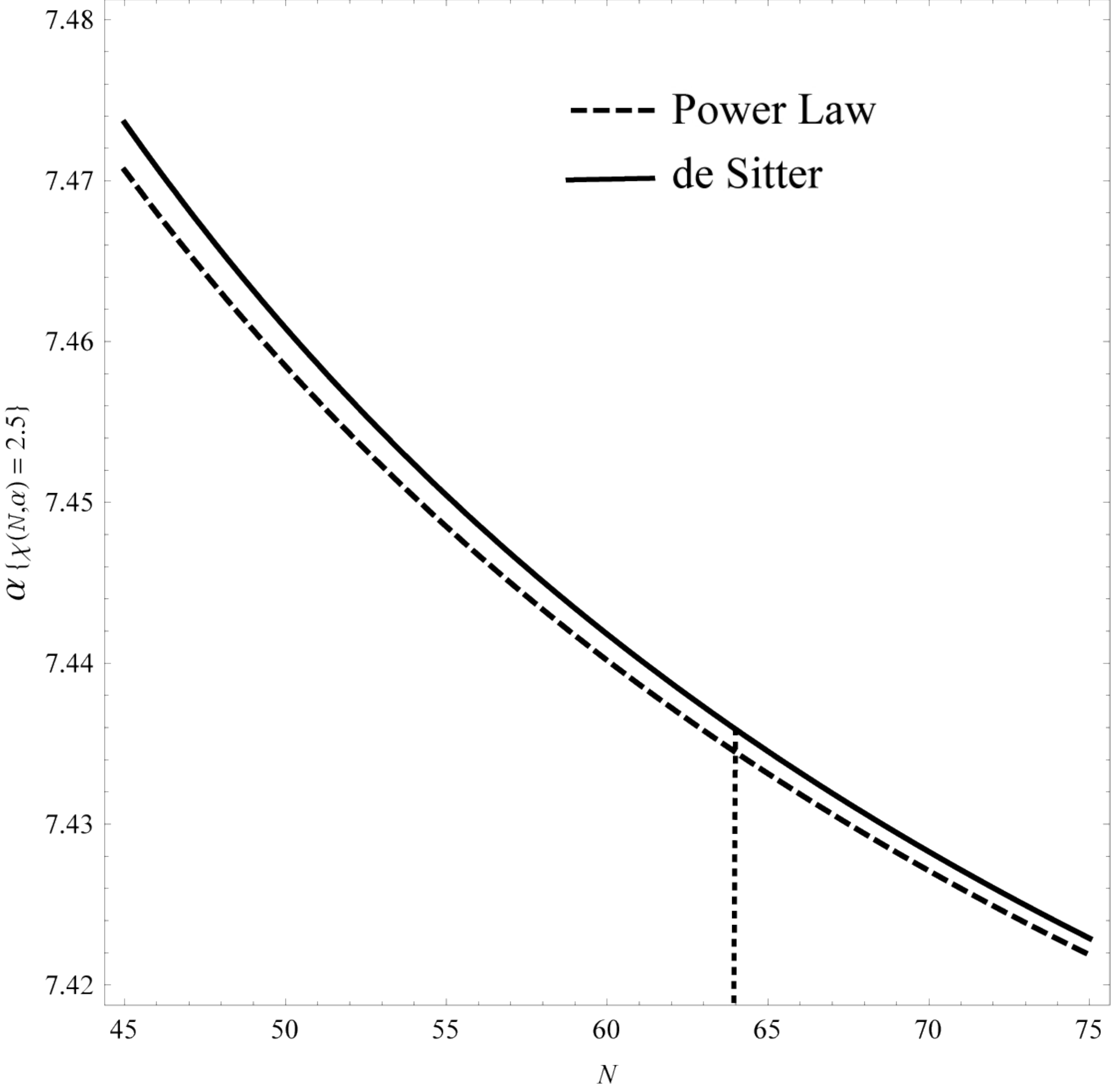}

\caption{\label{fig:5} The relation between $\alpha $ and $N$ at which $\chi (N,\alpha ) = 5/2$, for both de Sitter and power law expansion. For the interesting range of $N{\rm{ (50 < }}N{\rm{ < 70)}}$, the value of $\alpha$ is around, $\alpha  \sim {7.44.}$.}
\end{center}
\end{figure}

The value of the index, $\alpha $, at which the scale invariant PMF can be generated, varies as the value of $N$ changes. There is a slit different between this relation in de Sitter and power law expansion, see Fig.\ref{fig:5}. These values are around, $\alpha  \sim 7.44$, for an interesting values of e-folds, ($50 < N < 70$). The validity of $\alpha  \sim 7.44$ will be discussed in the section. \ref{sec:summary}, but from now onward, we adopt this value at $N = 64$ to study the scale invariant PMF. In general, the whole results of de Sitter are very close to the power law results but both of them will be presented for a completeness.

\begin{figure}[tbp]
\centering  \begin{center}
\includegraphics[width=0.85\textwidth,origin=c]{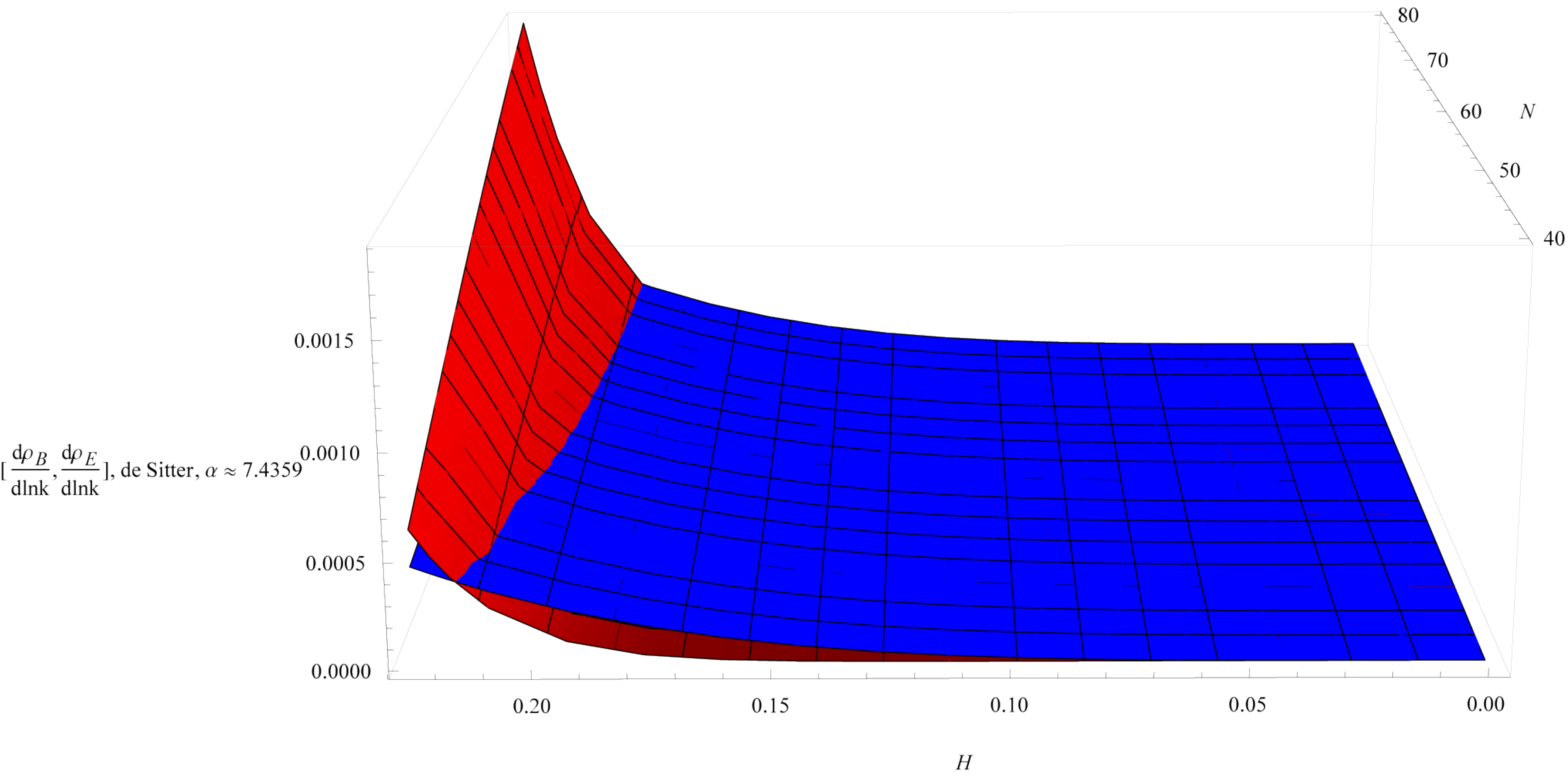}

\caption{\label{fig:6} The relation between the EM spectra, Hubble parameter $H$, and e-folds number, $N$ for de Sitter expansion. Around $H\sim 0.2{M_{{\rm{Pl}}}}$, electric field can exceed the magnetic field in all interesting range of $N{\rm{ (50 < }}N{\rm{ < 70)}}$. Hence, one can consider $H \sim 0.2{M_{{\rm{Pl}}}}$, as an upper limit under which the backreaction problem can be avoided.}
\end{center}
\end{figure}

First, let us draw the relation between electromagnetic spectra, Hubble parameter, $H$, and e-folding number, $N$, see Fig.\ref{fig:6}. It is clear, that, for the values of $H \le 0.2{M_{{\rm{Pl}}}}$, the backreaction problem may be avoided for all interesting values of $N$. Therefore, one can consider $H \sim 0.2{M_{{\rm{Pl}}}}$, as an upper limit of expansion rate below which, the backreaction problem can be avoided. The current investigation will be bounded in ${10^{ - 5}}{M_{{\rm{Pl}}}} < H < 0.2{M_{{\rm{Pl}}}}$. 

\begin{figure}[tbp]
\centering  \begin{center}
\includegraphics[width=0.8\textwidth,origin=c]{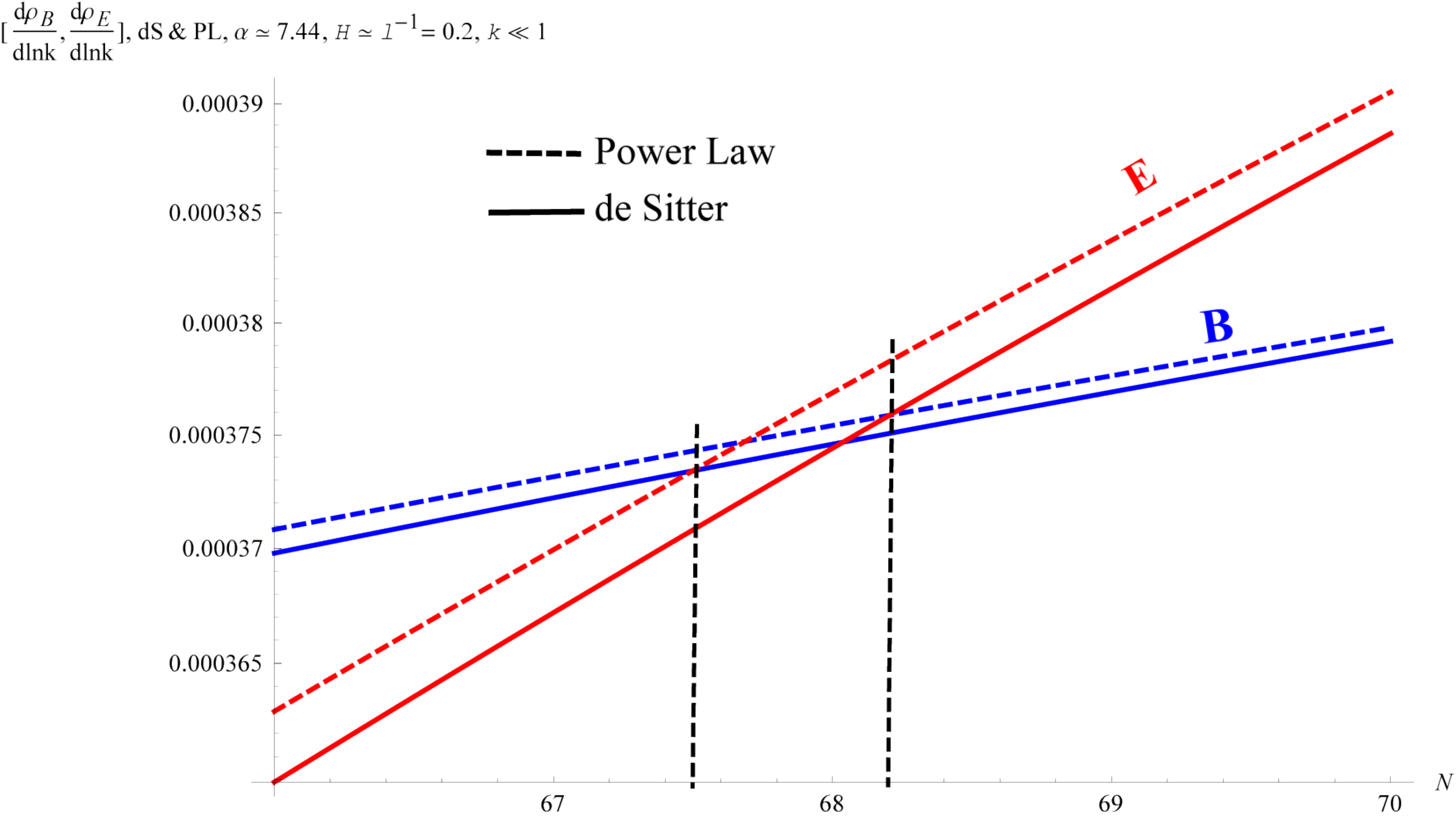}

\caption{\label{fig:7} The relation between the EM spectra and e-folds number, $N$ for both de Sitter and power law expansion around the value $H \sim 0.2{M_{{\rm{Pl}}}}$. The electric field can exceed the magnetic field around $N \sim 68$. Hence, one can consider it as an upper limit of e-folds number before the end of inflation under which the backreaction problem can be avoided.}
\end{center}
\end{figure}

Similarly, drawing the relation between the electromagnetic spectra and e-folding number around the upper limit, $H \sim 0.2{M_{{\rm{Pl}}}}$, shows that below the value, $N \sim 68$, the backreaction problem can be avoided, see Fig.\ref{fig:7}. Therefore, this value of $N \sim 68$, can be considered as an upper limit below which, a scale invariant PMF does suffer from the backreaction problem.

\subsection{Inflationary electromagnetic spectra in power law expansion}
\label{subsec:EM in PL} 

For more optimal slow roll analysis that has a smooth exit from inflation, one can use the power law expansion at which the Hubble parameter is not constant but a function of field, $H(\phi )$. If the field falls below a certain value, it starts to oscillate and then converts into particles in the reheating era, right after inflation. The scale factor follows the power law, (\ref{eq:12}). In the slow roll limit, we can approximate . Hence, substituting (\ref{eq:6}) into (\ref{eq:12}) yields,

\begin{equation}
\label{eq:42}
a(\eta ) = l{\rm{ }}{\eta ^{ - 1 - \frac{4}{{3{{\left( { - 1 + \frac{{4N}}{3}} \right)}^2}}}}},
\end{equation}
where, $l$ is the integration constant. For relatively high ($N \ge 50$), the absolute value of the exponent in (\ref{eq:42}) is  $\lesssim 1.0003$, so we can approximate $l \simeq {H^{ - 1}}$. Therefore, one cannot expect significant differences between the two cases. Nevertheless, we will investigate this case to have more precise results. 

\begin{figure}[tbp]
\centering  \begin{center}
\includegraphics[width=0.7\textwidth,origin=c]{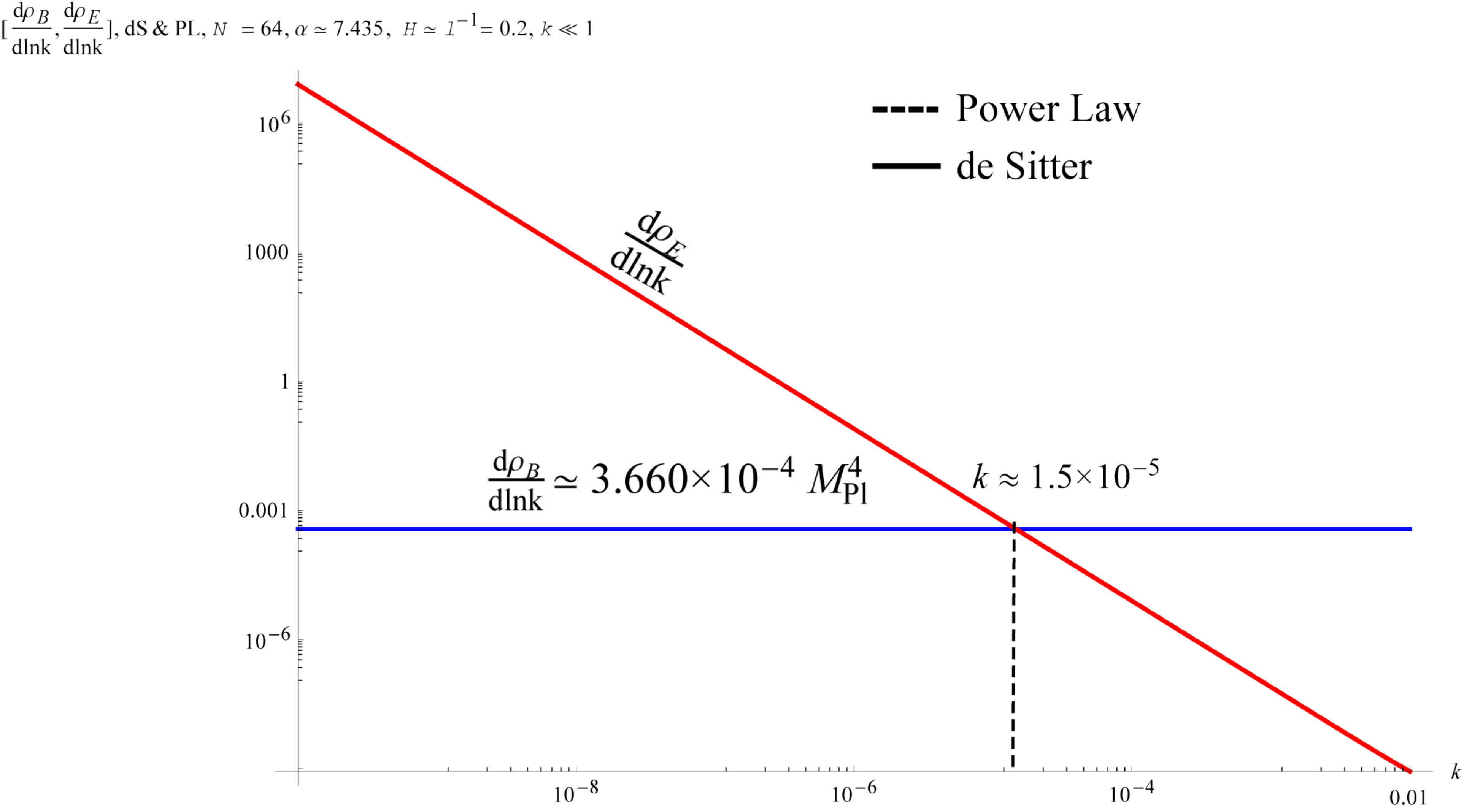}

\caption{\label{fig:8} The EM spectra for $\alpha  = {\rm{7}}{\rm{.4345}}$, $N = 64$, $\eta  =  - {10^{ - 2}}$ and the expansion rate, $H \approx {l^{ - 1}} = 0.2{M_{{\rm{Pl}}}}$. The PMF exceeds the electric fields at $k \sim 1.5 \times {10^{ - 5}}$.}
\end{center}
\end{figure}

\begin{figure}[tbp]
\centering  \begin{center}
\includegraphics[width=0.7\textwidth,origin=c]{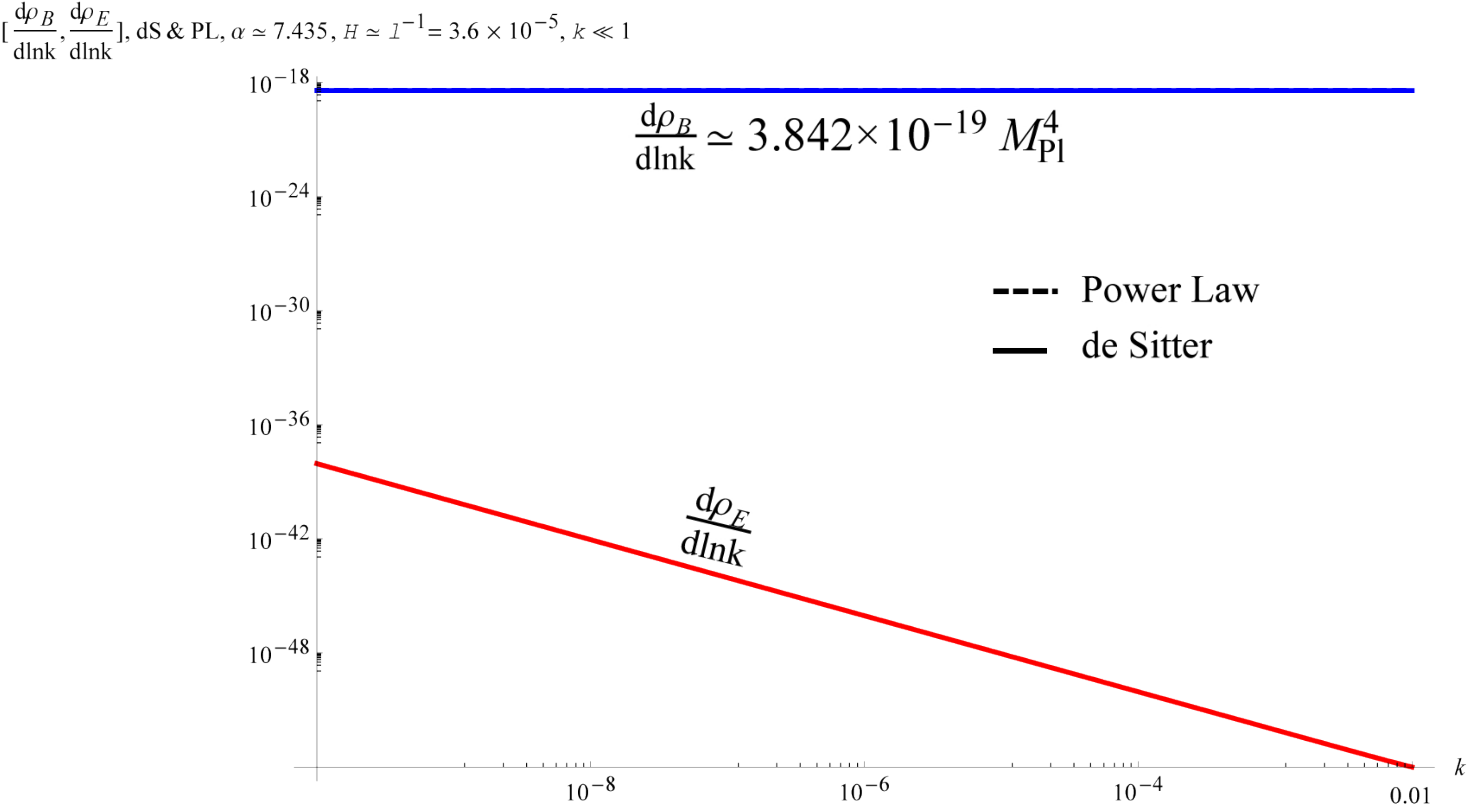}

\caption{\label{fig:9} The EM spectra for $\alpha  = {\rm{7}}{\rm{.4345}}$, $N = 64$, $\eta  =  - {10^{ - 2}}$ and the expansion rate at the onset of inflation, $H \approx {l^{ - 1}} = 3.6 \times {10^{ - 5}}{M_{{\rm{Pl}}}}$. The relation, $\frac{{d{\rho _B}}}{{d\ln k}} \gg \frac{{d{\rho _E}}}{{d\ln k}}$ stays valid, throughout inflation at low values of, $k \ll 1$.}
\end{center}
\end{figure}

Again, as we did in the last section, we will assume that $N$ is quasi-constant and we will not write it as a function of $\eta$ explicitly. Hence in the power law expansion, we substitute (\ref{eq:12}) into (\ref{eq:20}) and differentiating both sides, yields that,
\begin{equation}
\label{eq:43}
{Y_{PL}}\left( \eta  \right) = \frac{{{{\left( {3 - 4N} \right)}^2}{{\left( {63 - 4N\left( {63 + 4N\left( { - 9 + 4N} \right)} \right)} \right)}^2}\alpha \left( { - 36 + {{\left( {3 - 4N} \right)}^2}\alpha } \right)}}{{144{N^2}{{\left( {21 + 8N\left( { - 3 + 2N} \right)} \right)}^4}{\eta ^2}}},
\end{equation}

\begin{equation}
\label{eq:44}
\begin{split}
{\chi _{PL}}(\alpha ,N) = \frac{{\sqrt {\left( \begin{array}{l}
36{N^2}{\left( {21 + 8N\left( { - 3 + 2N} \right)} \right)^4} - 36{\left( {3 - 4N} \right)^2}{\left( {63 - 4N\left( {63 + 4N\left( { - 9 + 4N} \right)} \right)} \right)^2}\alpha \\
 + {\left( {3 - 4N} \right)^4}{\left( {63 - 4N\left( {63 + 4N\left( { - 9 + 4N} \right)} \right)} \right)^2}{\alpha ^2}
\end{array} \right))} }}{{12N{{\left( {21 + 8N\left( { - 3 + 2N} \right)} \right)}^2}}}
\end{split}
\end{equation}
Plugging (\ref{eq:43}) into (\ref{eq:36}) yields the same solution as (\ref{eq:38}). In this case, ${\chi _{PL}}(\alpha ,N) = 5/2$, at $\alpha  \simeq \{  - 7.4339,{\rm{ }}7.4345\} $, see Fig.\ref{fig:2} for all interesting values of N. Therefore, the scale invariant PMF can be obtained in the long wavelength regime, $k \ll 1$ (outside Hubble radius). Also, the electromagnetic spectra can be obtained in the same way as done in last section. The results is very close to the those obtained by de Sitter case, see Fig.\ref{fig:8}-\ref{fig:9}.

\begin{figure}[tbp]
\centering  \begin{center}
\includegraphics[width=0.7\textwidth,origin=c]{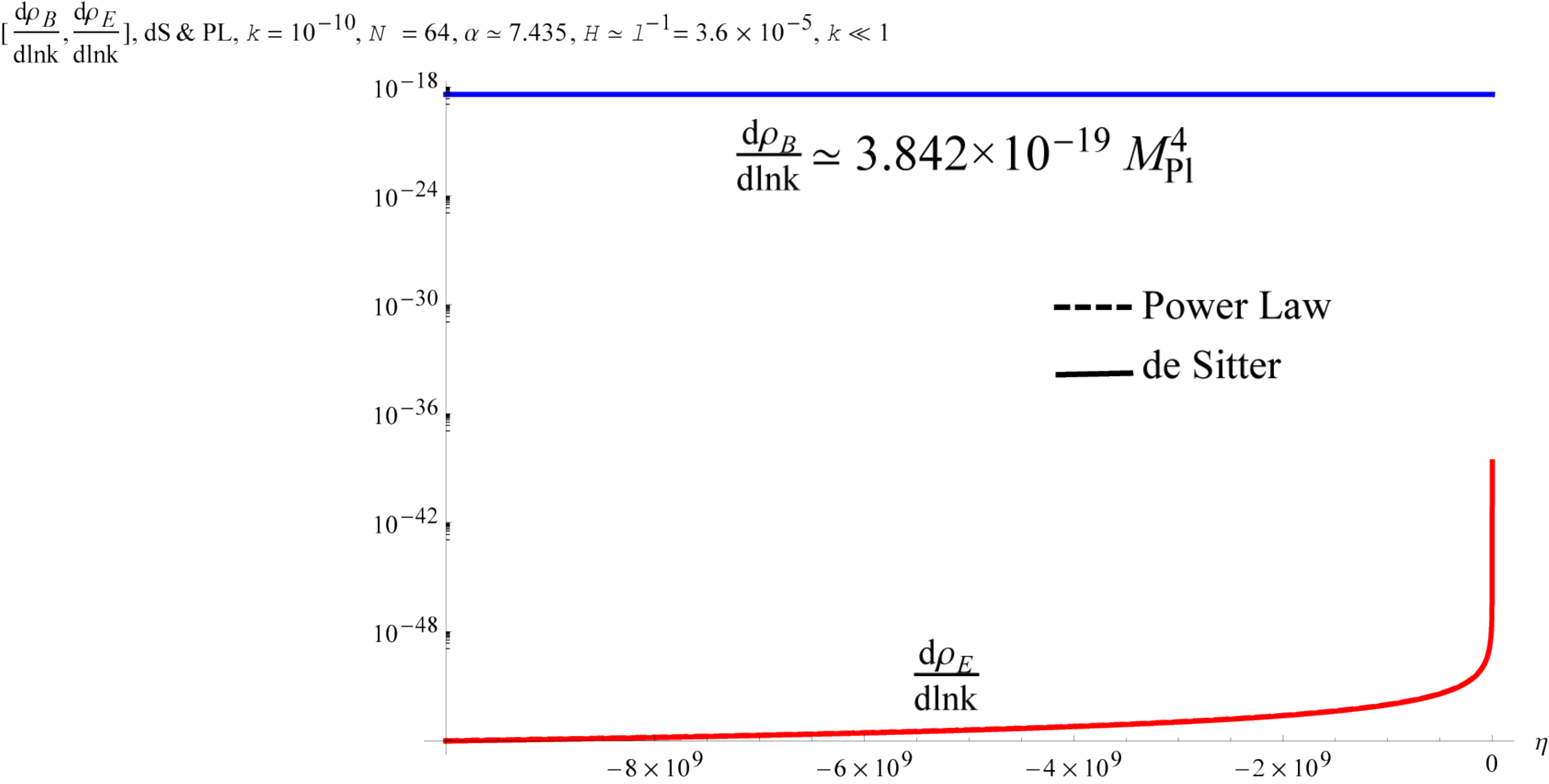}

\caption{\label{fig:10} The EM spectra for $\alpha  = {\rm{7}}{\rm{.4345}}$, $N = 64$, $k = {10^{ - 10}}$ and the upper limit of Hubble parameter during inflation, $H \approx {l^{ - 1}} = 3.6 \times {10^{ - 5}}$. The relation, $\frac{{d{\rho _B}}}{{d\ln k}} \gg \frac{{d{\rho _E}}}{{d\ln k}}$ stays valid, throughout the period, $\eta  \ll  - 1$ .}
\end{center}
\end{figure}

It is clear from Fig.\ref{fig:9} that, $\frac{{d{\rho _B}}}{{d\ln k}} \gg \frac{{d{\rho _E}}}{{d\ln k}}$ at $H \simeq {H_*} = 3.6 \times {10^{ - 5}}{M_{{\rm{Pl}}}}$. In fact that value of the Hubble parameter is the upper value during inflation or at the time of pivot scale, at which the space time exits the Hubble radius \cite{37}. Therefore, the problem of the backreaction caused by the divergence of the electric field at $k \ll 1$, is avoided. In order to make sure, the above relation stays valid throughout the inflationary era, we can plot the EM spectra versus the conformal time, at $\eta  \ll  - 1$, see Fig.\ref{fig:10}.

So far, we make sure that the electric field has negligible contribution during the inflation. However, to test the contribution of the magnetic field in the inflationary energy, one has to consider the constraints of post-inflation eras. That point will be discussed in the next section.

\section{Constraining reheating parameters by PMF}
\label{sec:Reh by PMF}

It is widely believed, that there was a phase of pre-heating or reheating at the end of inflation and before the radiation dominated era \cite{47}-\cite{49}. In this phase, as the temperature of inflation falls to certain value, ${T_{{\rm{reh}}}}$, inflaton starts oscillates around some value and decays into standard matter, that populates the Universe later. As the temperature continues to fall, the Big Bang Nucleosynthesis (BBN) starts taking place at ${T_{BBN}} \sim 1{\rm{MeV}}$. These produced particles are perturbatively decayed into radiation in the radiation era. 

Initially, this process was thought it only occurs in a complicated inflationary model that has more than one field \cite{50}-\cite{52}. However, later on it was shown that it could be occur even in a single field model but at sub-Hubble scale of perturbations \cite{53}-\cite{55}. In order to constrain this phase, one has to define the reheating parameters based on both the inflationary model and the observations, like the Cosmic Microwaves Background (CMB) \cite{56}, the Large Scale Structure (LSS) \cite{57}, the BBN \cite{58}, and magnetogenesis \cite{59}.

There are two main difficulties in investigating this era. First, no direct cosmological observations can constrains the reheating parameters. Second, the physics of this period is highly uncertain. Therefore, indirect constraints of the reheating parameters are usually calculated from other cosmological observations. Also, several models have been proposed to explain this era. In this section, the effect of the scale invariant PMF on the parameters of reheating is investigated. That is basically similar to Ref \cite{59} but by considering ${R^2}$-inflationary model. We adopt the new upper limits of present PMF which was constrained by Planck, 2015 \cite{9} and the instantaneous transition to the reheating and the epochs come after. For this reason, we start by discussing the reheating parameters. Next, Next, we discuss how to constrain them by ${R^2}$-inflation and the present upper limit of PMF.

\subsection{Reheating analysis}
\label{subsec:Reh Analys}

The reheating era can be specified mainly by three parameters, the reheating parameter, ${R_{{\rm{rad}}}}$, the reheating temperature, ${T_{{\rm{reh}}}}$, and the equation of state parameter, ${w _{{\rm{reh}}}}$ \cite{43},\cite{60}-\cite{61}. The first one is defined as,
\begin{equation}
\label{eq:45}
{R_{{\rm{rad}}}} \equiv \frac{{{a_{{\rm{end}}}}}}{{{a_{{\rm{reh}}}}}}{\left( {\frac{{{\rho _{{\rm{end}}}}}}{{{\rho _{{\rm{reh}}}}}}} \right)^{1/4}},
\end{equation}
where, $\rho$ is the energy density, and  \textquotedblleft end\textquotedblright and \textquotedblleft reh\textquotedblright stand respectively to the end of inflation and the end of reheating era. From the conservation of energy during the reheating era, $\rho  = {\rho _\phi } + {\rho _\gamma }$ and $P = {P_\phi } + {\rho _\gamma }/3$, where ${\rho _\phi }$ is the energy density of the scalar field of inflation and ${\rho _\gamma }$ is the energy density of the radiation. Assuming these are the main constituent of the reheating era, one can write,
\begin{equation}
\label{eq:46}
{\rho _{{\rm{reh}}}} = {\rho _{{\rm{end}}}}\exp \left\{ { - 3\int_{{N_{{\rm{end}}}}}^{{N_{{\rm{reh}}}}} {\frac{{\left[ {1 + w (a)} \right]}}{a}da} } \right\} = {\rho _{{\rm{end}}}}\exp [ - 3\Delta N(1 + {\bar w _{{\rm{reh}}}})],
\end{equation}
where, ${\bar w _{{\rm{reh}}}}$ is mean value of the equation of state parameter ($P/\rho  = w(a)$) defined by,
\begin{equation}
\label{eq:47}
{\bar w _{{\rm{reh}}}} \equiv \frac{1}{{{N_{{\rm{reh}}}} - {N_{{\rm{end}}}}}}\int_{{N_{{\rm{end}}}}}^{{N_{{\rm{reh}}}}} {\frac{{w (a)}}{a}da}.
\end{equation}
Taking the logarithm of (\ref{eq:45}) and making use of (\ref{eq:46})yields, 
\begin{equation}
\label{eq:48}
\ln {R_{{\rm{rad}}}} = \frac{{\left( {{N_{{\rm{reh}}}} - {N_{{\rm{end}}}}} \right)}}{4}\left( { - 1 + 3{{\bar w }_{{\rm{reh}}}}} \right) = \frac{{1 - 3{{\bar w }_{{\rm{reh}}}}}}{{12(1 + {{\bar w }_{{\rm{reh}}}})}}\ln \left( {\frac{{{\rho _{{\rm{reh}}}}}}{{{\rho _{{\rm{end}}}}}}} \right),
\end{equation}
In terms of the pivot quantities \cite{30},
\begin{equation}
\label{eq:48.5}
\ln {R_{{\rm{rad}}}} = {N_ * } + \ln \left( {\frac{{k/{a_0}}}{{\rho _\gamma ^{1/4}}}} \right) + \frac{1}{4}\ln \left( {\frac{{3{V_{{\rm{end}}}}}}{{{\epsilon _{1V * }}{V_ * }}}\frac{{3 - {\epsilon _{1V * }}}}{{3 - {\epsilon _{1V{\rm{end}}}}}}} \right) - \frac{1}{4}\ln \left( {\frac{{H_ * ^2}}{{M_{{\rm{Pl}}}^2{\epsilon _{1V * }}}}} \right),
\end{equation}

On the other hand, the relation between the reheating temperature and energy density can be written as,
\begin{equation}
\label{eq:49}
{\rho _{{\rm{reh}}}} = \frac{{{\pi ^2}}}{{30}}{g_{{\rm{reh}}}}T_{{\rm{reh}}}^4,
\end{equation}
where, ${g_{{\rm{reh}}}}$ is the number of relativistic degree of freedom at the end of reheating. One also can relate the reheating temperature to today temperature of CMB, ${T_{\rm{0}}}$, \cite{62} as
\begin{equation}
\label{eq:50}
{T_{{\rm{reh}}}} = {T_0}\left( {\frac{{{a_0}}}{{{a_{{\rm{reh}}}}}}} \right){\left( {\frac{{43}}{{11{g_{{\rm{reh}}}}}}} \right)^{1/3}} = {T_0}\left( {\frac{{{a_0}}}{{{a_{{\rm{eq}}}}}}} \right){e^{{N_{{\rm{eq}}}}}}{\left( {\frac{{43}}{{11{g_{{\rm{reh}}}}}}} \right)^{1/3}}.
\end{equation}
where, ${a_{{\rm{eq}}}}/{a_{{\rm{reh}}}} = {e^{{N_{{\rm{eq}}}}}}$, during radiation era, and \textquotedblleft eq\textquotedblright  stands for the period of equality between radiation and matter dominant phases.  	

In terms of the pivot scale, $k = {a_k}{H_k}$, at which the commoving scale crosses the Hubble radius (horizon size) during the inflation, one can write down the relation between the scale factors and Hubble parameters \cite{43} for different epochs of the Universe as
\begin{equation}
\label{eq:51}
\frac{k}{{{a_0}{H_0}}} = \frac{{{a_k}{H_k}}}{{{a_0}{H_0}}} = \frac{{{a_k}}}{{{a_{{\rm{end}}}}}}\frac{{{a_{{\rm{end}}}}}}{{{a_{{\rm{reh}}}}}}\frac{{{a_{{\rm{reh}}}}}}{{{a_{{\rm{eq}}}}}}\frac{{{a_{{\rm{eq}}}}}}{{{a_0}}}\frac{{{H_k}}}{{{H_0}}}.
\end{equation}
From (\ref{eq:51}), one can write the ration,
\begin{equation}
\label{eq:52}
\frac{{{a_0}}}{{{a_{{\rm{eq}}}}}} = \frac{{{a_{\rm{0}}}{H_k}}}{k}{e^{ - {N_k}}}{e^{ - {N_{{\rm{reh}}}}}}{e^{ - {N_{{\rm{eq}}}}}}.
\end{equation}
In (\ref{eq:50}) and (\ref{eq:52}), we assume a form of exponential expansion takes place during reheating and radiation epoch too. Substituting of (\ref{eq:52}) into (\ref{eq:50}) gives,
\begin{equation}
\label{eq:53}
{T_{{\rm{reh}}}} = \left( {\frac{{{T_0}{a_0}}}{k}} \right){\left( {\frac{{43}}{{11{g_{{\rm{reh}}}}}}} \right)^{1/3}}{H_k}{e^{ - {N_k}}}{e^{ - {N_{{\rm{reh}}}}}}.
\end{equation}

If the equation of state is assumed to be constant during reheating then, $\rho  \propto {a^{ - 3(1 + w )}}$. Also, the relation between the energy density and the potential \cite{63} can be written as,
\begin{equation}
\label{eq:54}
\rho  = \left( {1 + \frac{1}{{3/{\epsilon _{1V}} - 1}}} \right)V\left(\phi\right).
\end{equation}
Making use of these relations and substituting of (\ref{eq:53}) into (\ref{eq:49}) and taking natural logarithm yields the equations of reheating e-folds numbers and temperature,
\begin{equation}
\label{eq:55}
\begin{split}
{N_{{\rm{reh}}}} = \frac{4}{{1 - 3{w _{{\rm{reh}}}}}}\left[ \begin{array}{l}
 - {N_k} - \ln \left( {\frac{k}{{{a_0}{T_0}}}} \right) - \frac{1}{4}\ln \left( {\frac{{30}}{{{g_{{\rm{reh}}}}{\pi ^2}}}} \right) - \frac{1}{3}\ln \left( {\frac{{11{g_{{\rm{reh}}}}}}{{43}}} \right)\\
 - \frac{1}{4}\ln ({V_{{\rm{end}}}}) - \frac{1}{4}\ln \left( {1 + \frac{1}{{3/{\epsilon _{1V}} - 1}}} \right) + \frac{1}{2}\ln \left( {\frac{{{\pi ^2}r{A_s}}}{2}} \right)
\end{array} \right],
\end{split}
\end{equation}

\begin{equation}
\label{eq:56}
{T_{{\rm{reh}}}} = {\left[ {\left( {\frac{{{T_0}{a_0}}}{k}} \right){{\left( {\frac{{43}}{{11{g_{{\rm{reh}}}}}}} \right)}^{1/3}}{H_k}{e^{ - {N_k}}}{{\left[ {\left( {1 + \frac{1}{{3/{\epsilon _{1V}} - 1}}} \right)\frac{{{V_{{\rm{end}}}}}}{{{\pi ^2}{g_{{\rm{reh}}}}}}} \right]}^{ - \frac{1}{{3(1 + {w _{{\rm{reh}}}})}}}}} \right]^{\frac{{3(1 + {w _{{\rm{reh}}}})}}{{3{w _{{\rm{reh}}}} - 1}}}},
\end{equation}
where, ${V_{{\rm{end}}}}$ is the potential at the end of inflation and ${A_s}$ is the scalar power spectrum magnitude, obtained by (\ref{eq:16}). Hence, from (\ref{eq:2}) and the condition, ${\epsilon _{nV}} \approx 1$, at the end of inflation, one can decide ${V_{{\rm{end}}}}$. In fact, the field ${\phi _{{\rm{end}}}}$ has three different values, based on which slow roll parameter is equal to unity. They are respectively, ${\phi _{{\rm{end}}}} \simeq {\rm{0}}{\rm{.94}}{M_{{\rm{Pl}}}},{\rm{ 1}}{\rm{.83}}{M_{{\rm{Pl}}}},{\rm{ 1}}{\rm{.51}}{M_{{\rm{Pl}}}}$ for  ${\epsilon _{1V}},{\rm{ }}{\epsilon _{2V}}$ and ${\epsilon _{3V}}$. Therefore, we will pick the first value to avoid violating the slow roll condition. 

By adopting Planck, 2015 results \cite{37}, \cite{62}-\cite{63}, we have $\frac{k}{{{a_0}}} = 0.05{\rm{Mp}}{{\rm{c}}^{ - 1}}$, ${n_s} = {\rm{ 0}}{\rm{.9682}} \pm 0.0062$, ${A_s} = {\rm{ 2}}{\rm{.196}} \times {\rm{1}}{{\rm{0}}^{ - 9}}$, and ${H_k} < 3.6 \times {10^{ - 5}}{M_{{\rm{Pl}}}}$, also by using ${g_{{\rm{reh}}}} \approx 100$,  Eqs.(\ref{eq:55})-(\ref{eq:56}) become
\begin{equation}
\label{eq:57}
{N_{{\rm{reh}}}} = \frac{4}{{1 - 3{w _{{\rm{reh}}}}}}\left[ {61.1 - {N_k} - \ln \left( {\frac{{V_{{\rm{end}}}^{1/4}}}{{{H_k}}}} \right)} \right],
\end{equation}
	
\begin{equation}
\label{eq:58}
{T_{{\rm{reh}}}} = \exp \left[ { - \frac{3}{4}(1 + {w _{{\rm{reh}}}}){N_{{\rm{reh}}}}} \right]{\left[ {\left( {\frac{3}{{10{\pi ^2}}}} \right)\left( {1 + \frac{1}{{3/{\epsilon _{1V}} - 1}}} \right){V_{{\rm{end}}}}} \right]^{1/4}}.
\end{equation}	
The equation of state parameter at reheating era, ${w _{{\rm{reh}}}}$, for general inflationary potential is usually taken in the interval, $ - 1/3 < {w _{{\rm{reh}}}} < 1$ \cite{13}. But for Starobinsky inflation (${R^2}$-inflation), the interval that fits well with Planck, 2015 results is  $0 < {w _{{\rm{reh}}}} < 1/3$ \cite{62}.

\subsection{Constraining reheating parameters by the present PMF in ${R^2}$-inflation}
\label{subsec: Reh par by PMF}

The next step is to relate the reheating parameters to the present PMF which is constrained by the results of Planck, 2015 \cite{9}. As the conformal invariance of electromagnetic field is restored after inflation, the present super Hubble magnetic field ${B_0}$ is redshifted since the end of inflation \cite{59} as,

\begin{equation}
\label{eq:59}
{B_0} = \frac{{{B_{{\rm{end}}}}}}{{{{(1 + {z_{{\rm{end}}}})}^2}}},
\end{equation}
where, ${B_{{\rm{end}}}}$ is the magnetic field at the end of inflation and ${z_{{\rm{end}}}}$ is the redshift at the end of inflation. Hence, at the end of inflation we can write,

\begin{equation}
\label{eq:60}
1 + {z_{{\rm{end}}}} = \frac{{{a_0}}}{{{a_{{\rm{end}}}}}} = \frac{{{a_0}}}{{{a_{{\rm{eq}}}}}}\frac{{{a_{{\rm{eq}}}}}}{{{a_{{\rm{reh}}}}}}\frac{{{a_{{\rm{reh}}}}}}{{{a_{{\rm{end}}}}}} = \frac{1}{{{R_{{\rm{rad}}}}}}{\left( {\frac{{{\rho _{{\rm{end}}}}}}{{{\rho _\gamma }}}} \right)^{1/4}},
\end{equation}
where,${\rho _\gamma }\sim (5.7 \times {10^{ - 125}}M_{Pl}^4)$ is energy density of  radiation today.

We substitute (\ref{eq:60}) into (\ref{eq:59}) and make sure there is no backreaction problem at the end of inflation, ${\rho _{{B_{{\rm{end}}}}}} < {\rho _{{\rm{end}}}}$. Since, ${\rho _B} = {B^2}/2$, then combining (\ref{eq:59}) and (\ref{eq:60}) yields the constraint in reheating parameters from the present PMF \cite{59},
	
\begin{equation}
\label{eq:61}
{R_{{\rm{rad}}}} \gg \frac{{B_0^{1/2}}}{{{{(2{\rho _\gamma })}^{1/4}}}}.
\end{equation}
The upper limit of present PMF calculated by Planck, 2015 is $\sim 10^{ - 9}\rm{G}$. Therefore, the lower limit of reheating parameter is, $R_{\rm{rad}} > 1.761 \times 10^{- 2}$. But this limit is independent of the inflationary model. Hence, for a more accurate limit of $R_{\rm{rad}}$ associated with $R^{2}$- inflation, one can use the result of \cite{49} and adopting the Planck, 2015 results, at which the middle value, ${N_\ast} \simeq 58.5$, if $w _{\rm{reh}}$ is not constant \cite{37}. It implies that, $R_{\rm{rad}} > 6.888$, which is three orders of magnitude more than the previous one. Therefore, the reheating in this case is more constraint by the inflationary model bound. 

In order to constrain the reheating energy scale, one can substitute (\ref{eq:61}) into (\ref{eq:48}) to obtain,
	
\begin{equation}
\label{eq:62}
{\rho _{{\rm{reh}}}} > {\rho _{{\rm{end}}}}{\left[ {\frac{{{B_0}}}{{{{(2{\rho _\gamma })}^{1/2}}}}} \right]^{\frac{{6(1 + {{\bar w }_{{\rm{reh}}}})}}{{1 - 3{{\bar w }_{{\rm{reh}}}}}}}}.
\end{equation}
Therefore, one has to find $\rho _{\rm{end}}$ for the model of inflation. Also, the lower limit of reheating energy density should be greater than BBN energy, $\rho _{\rm{reh}}> \rho _{\rm{nuc}}$, where $\rho_{\rm{nuc}}$ is in the order of ($\sim 10{\rm{MeV}}$). 

We can use the upper bound of the energy density of inflation derived from WMAP7, ${\left( {{\rho _{{\rm{end}}}}} \right)_{{\rm{CMB}}}} < 2.789 \times {10^{ - 10}}M_{{\rm{Pl}}}^4$ \cite{56} and the lower limit of ${R_{{\rm{rad}}}}( > 6.888)$ and substitute them into (\ref{eq:60}), one can find the upper bound of, $1 + {z_{{\rm{end}}}} < 6.828 \times {10^{27}}$. If we substitute the redshift into \ref{eq:59}, we can find the upper limit of magnetic field at the end of inflation, ${B_{{\rm{end}}}} < 4.662 \times {10^{46}}{\rm{ G}}{\rm{(2}}.3541 \times {10^{ - 5}}M_{{\rm{Pl}}}^2)$, where we have used, $1{\rm{ G }} \approx {\rm{ 3}}{\rm{.3 }} \times {\rm{ 1}}{{\rm{0}}^{ - 57}}M_{{\rm{Pl}}}^2$. Hence, the energy density of the magnetic field is ${\left( {{\rho _{{B_{{\rm{end}}}}}}} \right)_{{\rm{CMB}}}} < 1.184 \times {10^{ - 20}}M_{{\rm{Pl}}}^4$. It is ten orders of magnitude more than the upper limit of ${\rho _{{\rm{end}}}}$ found from CMB \cite{56}. Therefore, the backreaction problem can be avoided easily. 

On the other hand, we can use Eq.(\ref{eq:54}) to find the energy density at the end of inflation $(\epsilon_{1V} \approx 1, \rho _{\rm{end}} = \frac{3}{2} V_{\rm{end}})$. By adopting the value of $\Lambda (\sim 4.0 \times 10^{-5} M_{\rm{Pl}})$, calculated from the amplitude of the CMB anisotropies \cite{30}, we find $\left( \rho_{\rm{end}} \right)_{{\rm{R}}^2-\rm{inflation}} \simeq 1.1 \times 10^{- 18}M_{\rm{Pl}}^4$. Similarly, substituting of this value with the limit, $R_{\rm{rad}}( > 6.888)$ into (\ref{eq:59})-(\ref{eq:60}) implies that, $1+z_{\rm{end}} < 5.414 \times {10^{25}}$ and $\left( \rho _{B_{\rm{end}}} \right)_{{{\rm{R}}^2}-\rm{inflation}} < 4.6788 \times 10^{ - 29}M_{\rm{Pl}}^4$, which is free more from backreaction problem. Therefore, $\left(\rho_{B_{\rm{end}}}\right)_{{\rm{R}}^2- \rm{inflation}}$ calculated by inflationary model and the present limits of PMF found by Planck, 2015 puts more constraints on the $\rho _{B_{\rm{end}}}$ than the constraints found by $\left( \rho_{B_{{\rm{end}}}}\right)_{\rm{CMB}}$ with present limits of PMF.
	
\begin{figure}[tbp]
\centering  \begin{center}
\includegraphics[width=0.7\textwidth,origin=c]{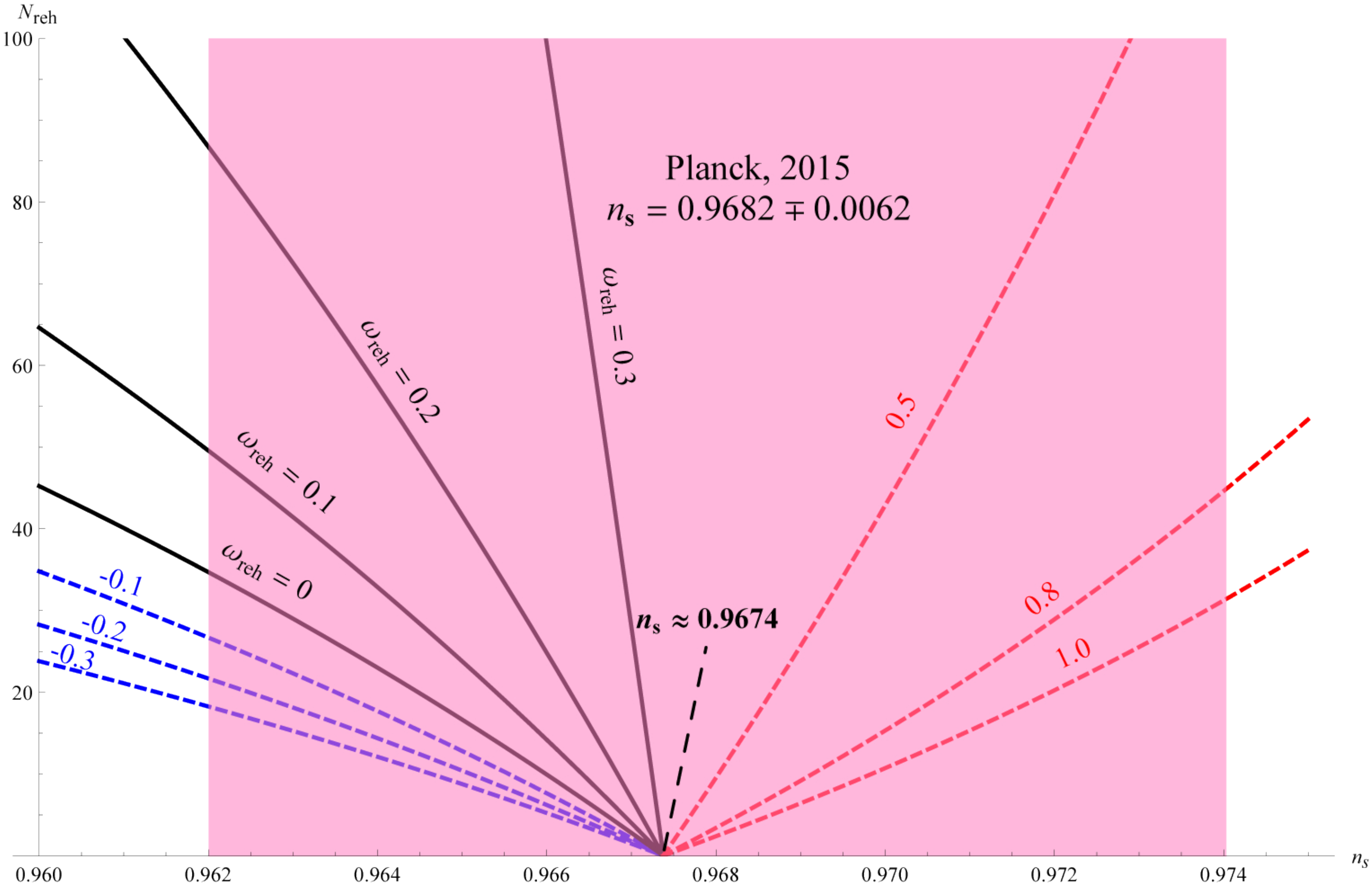}

\caption{\label{fig:11} The plot of reheating e-folds, ${N_{{\rm{reh}}}}$, versus spectral index, ${n_{\rm{s}}}$, at the end of ${R^2}$-inflation, for some values in $ - 0.3 < {w _{{\rm{reh}}}} < 1$. They all intersect at ${n_{\rm{s}}} \approx 0.9674$, which lies well in the Planck, 2015 range, ${n_s} = {\rm{ 0}}{\rm{.9682}} \pm 0.0062$}
\end{center}
\end{figure}

\begin{figure}[tbp]
\centering  \begin{center}
\includegraphics[width=0.7\textwidth,origin=c]{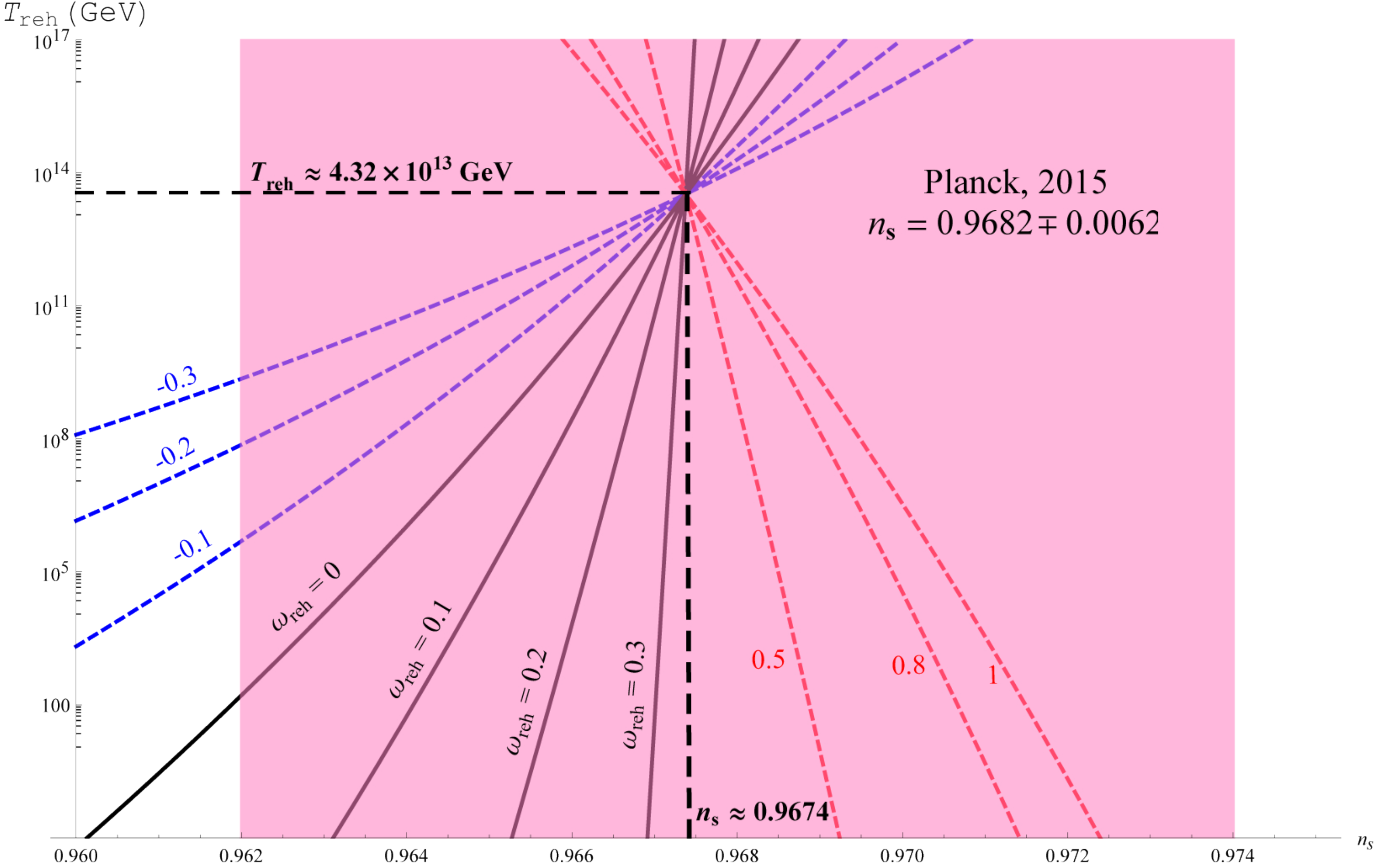}

\caption{\label{fig:12} The plot of reheating temperature, ${T_{{\rm{reh}}}}$, versus spectral index, ${n_{\rm{s}}}$, at the end of ${R^2}$-inflation, for some values in $ - 0.3 < {w _{{\rm{reh}}}} < 1$. They all intersect into ${T_{{\rm{reh}}}} \sim 4.32 \times {10^{13}}{\rm{GeV}}$ at ${n_{\rm{s}}} \approx 0.9674$. The value of the temperature is below ${({\rho _{{\rm{end}}}})^{1/4}}$ and the spectral index lie in the Planck, 2015 range, ${n_s} = {\rm{ 0}}{\rm{.9682}} \pm 0.0062$.}
\end{center}
\end{figure}

Similarly, one can plot both ${N_{{\rm{reh}}}}$ and ${T_{{\rm{reh}}}}$ as a function of ${n_{\rm{s}}}$ by using Eq.(\ref{eq:58}), see Fig.\ref{fig:11}-\ref{fig:12}. As we can see from Fig.\ref{fig:12}, all curves of possible ${w _{{\rm{reh}}}}$ intersect at ${T_{{\rm{reh}}}} \approx 4.32 \times {10^{13}}{\rm{GeV}}$ and ${n_{\rm{s}}} \approx 0.9674$, the value of spectral index fits well with the range of Planck, 2015 for ${R^2}$-inflation. This temperature is much more than the range of reheating temperature obtained from CMB in the context of LFI, see Eq.(\ref{eq:54}) of Ref.\cite{56}. As we use the upper limit of ${H_ * }$ in Eq.(\ref{eq:57}), we can consider the above value as the upper limit of ${T_{{\rm{reh}}}}$. 

\begin{figure}[tbp]
\centering  \begin{center}
\includegraphics[width=0.7\textwidth,origin=c]{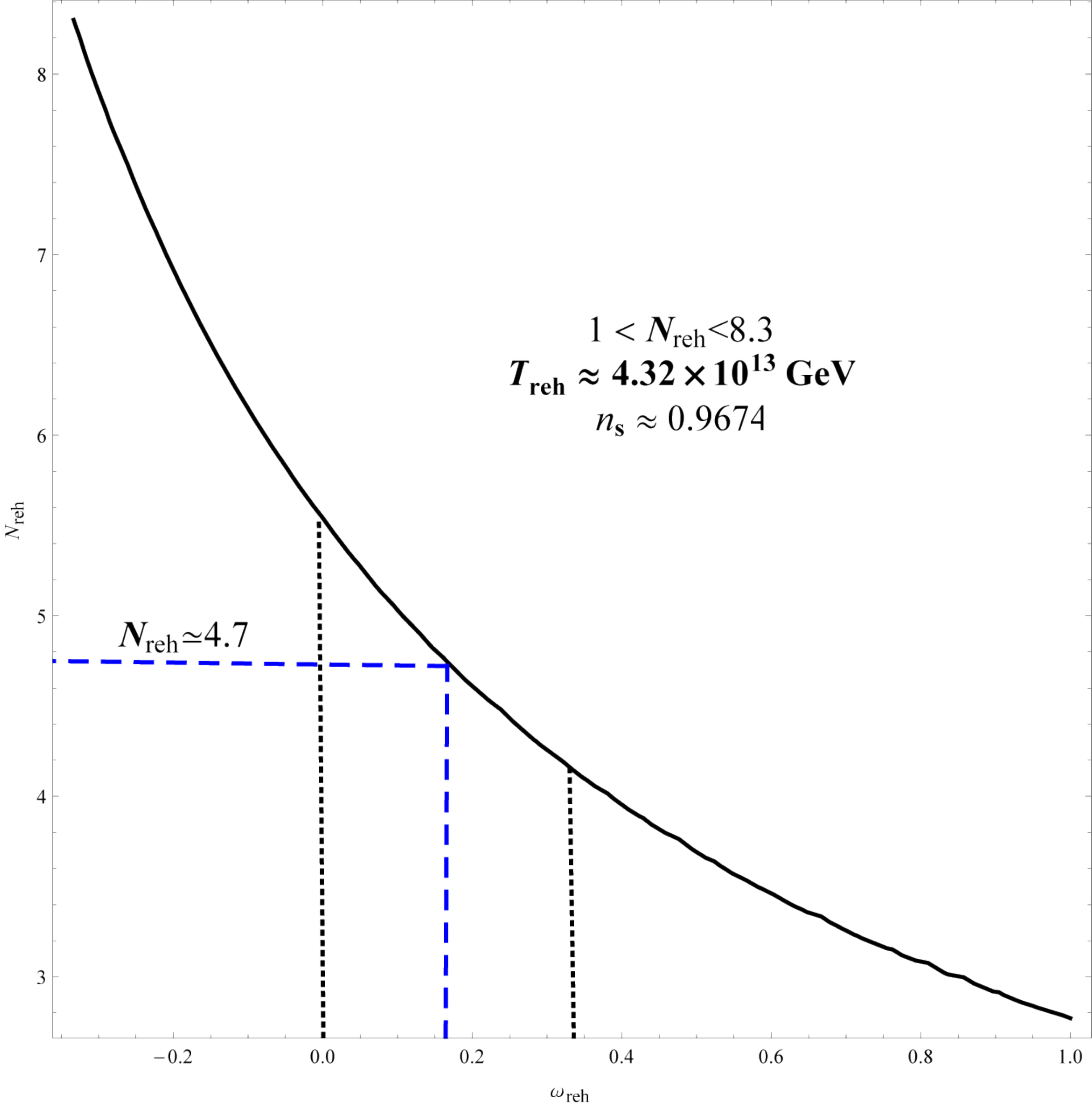}

\caption{\label{fig:13} The number of e-folds, ${N_{{\rm{reh}}}}$, during reheating, for all possible values of the equation of state parameter, $ - 1/3 < {w _{{\rm{reh}}}} < 1$, at ${n_{\rm{s}}} \approx 0.9637$. The range of e-folds is, $1 < {N_{{\rm{reh}}}} < 8.3$. In ${R^2}$-inflation, the values of state parameter lies onto $0 < {w _{{\rm{reh}}}} < 1/3$, hence the middle value is ${N_{{\rm{reh}}}} \simeq 4.7$.}
\end{center}
\end{figure}

	Thus, adopting the this temperature for all ${w _{{\rm{reh}}}}$ models of reheating, enables us to constrain the ${N_{{\rm{reh}}}}$ on the range $1 < {N_{{\rm{reh}}}} < 8.3$, for all possible values of ,${w _{{\rm{reh}}}}$ see Fig.\ref{fig:13}. The average value of ${N_{{\rm{reh}}}}$ is relatively low, ${N_{{\rm{reh}}}} \simeq 4.7$. Also, from, (\ref{eq:49}) one can find, ${\rho _{{\rm{reh}}}} \approx 3.259 \times {10^{ - 18}}M_{{\rm{Pl}}}^4$, which is in the same order of magnitude of the upper limit of ${\left( {{\rho _{{\rm{end}}}}} \right)_{{{\rm{R}}^2} - {\rm{inflation}}}}$ and two orders of magnitude less than the upper limit of ${\left( {{\rho _{{B_{{\rm{end}}}}}}} \right)_{{\rm{CMB}}}}$. Interestingly enough this value of $\rho _{{\rm{reh}}}^{1/4}(1.035 \times {10^{14}}{\rm{GeV}})$ is one order of magnitude less than the range of ${\rho _{{\rm{end}}}}$, obtained from CMB for the large field inflation \cite{56}, see Eq.(\ref{eq:45}) of the same reference. Therefore, the instantaneous reheating at which, ${\rho _{{\rm{end}}}} \approx {\rho _{{\rm{reh}}}}$, can be manifested by this result.
	
	At the end of this section, we constrain the upper limit of reheating energy density, ${\rho _{{\rm{reh}}}}$, based on the lower limits of both ${N_{{\rm{reh}}}}$ and reheating parameter, ${R_{{\rm{rad}}}}$. The last one was obtained from the present upper limit of PMF by Planck, 2015. Hence, Eq.(\ref{eq:45}) can be written as,
	
\begin{equation}
\label{eq:63}
{R_{{\rm{rad}}}} = {{\rm{e}}^{ - {N_{{\rm{reh}}}}}}{\left( {\frac{{{\rho _{{\rm{end}}}}}}{{{\rho _{{\rm{reh}}}}}}} \right)^{1/4}},
\end{equation}
If we use ${\left( {{\rho _{{\rm{end}}}}} \right)_{{\rm{CMB}}}}$, the upper limit, ${\left( {{\rho _{{\rm{reh}}}}} \right)_{{\rm{CMB}}}} < 8.480 \times {10^{ - 22}}M_{{\rm{Pl}}}^4$, which is much more than the lower limit derived from WMAP7 for both large and small field inflation \cite{56}. However, if ${\left( {{\rho _{{\rm{end}}}}} \right)_{{{\rm{R}}^2} - {\rm{inflation}}}}$ is used, the upper limit is ${\left( {{\rho _{{\rm{reh}}}}} \right)_{{{\rm{R}}^2} - {\rm{inflation}}}} < 3.344 \times {10^{ - 30}}M_{{\rm{Pl}}}^4$. It is still much more than the lower limit derived from WMAP7 for both large and small field inflation.

\section{Constraining the present PMF from the magnetogensis}
\label{sec:PMF from Magnetogens} 

In this section, we will constrain the value of present PMF, based on both the predicted scale invariant magnetic field generated in inflationary era by ${f^2}FF$model in ${R^2}$-inflation, which is discussed in section.\ref{sec:EM in R2}, and by using the constraint values of reheating parameters, which are calculated in section.\ref{sec:Reh by PMF}. In order to do that, we need to impose some necessary assumptions. First as shown in \cite{24}, the magnetic field energy density, ${\rho _{\rm{B}}}$, scales as $1/{a^4}$ independently of the dominant constituent in the reheating era and the eras come after. That is basically the implication of Eq.(\ref{eq:59}). Second, and to insure that, this model does not suffer from the backreaction problem, we generalize the validity of Eq.(102) in \cite{24} to the end of inflation.
\begin{equation}
\label{eq:64}
{\left. {\frac{{d{\rho _E}}}{{d\ln k}}} \right|_{{\rm{end}}}} + {\left. {\frac{{d{\rho _B}}}{{d\ln k}}} \right|_{{\rm{end}}}} < {\rho _{{\rm{end}}}},
\end{equation}
where, the EM spectra in (\ref{eq:64}) are calculated by (\ref{eq:32})-(\ref{eq:33}). The third assumption is that, the reheating is not going to affect the magnitude and the shape of EM spectra. 

As a result of the above assumptions, one can neglect the electric field in Fig.\ref{fig:9} without specifying the constraints of electric conductivity of the reheating (${\sigma _c} \gg H$) which may lead to zero electric field and constant magnetic field. Also, because of relatively small value of ${N_{{\rm{reh}}}}$, we can assume that, ${\left. {\frac{{d{\rho _B}}}{{d\ln k}}} \right|_{{\rm{end}}}} \simeq {\left. {\frac{{d{\rho _B}}}{{d\ln k}}} \right|_{{\rm{reh}}}}$. Since, we use Planck units in computation $({M_{{\rm{Pl}}}} = 1)$, then the scale of the spectra is in $(M_{{\rm{Pl}}}^4)$. By using the upper limit of Hubble parameter during inflation, $H = 3.6 \times {10^{ - 5}}{M_{{\rm{Pl}}}}$, the magnitude of magnetic spectra obtained from Fig.\ref{fig:9}-\ref{fig:10} is ${\left. {\frac{{d{\rho _B}}}{{d\ln k}}} \right|_{{\rm{end}}}} \simeq 3.842 \times {10^{ - 19}}M_{{\rm{Pl}}}^4$. This value is well below both the upper limit of ${\left( {{\rho _{{\rm{end}}}}} \right)_{{\rm{CMB}}}}$ and ${\left( {{\rho _{{\rm{end}}}}} \right)_{{{\rm{R}}^2} - {\rm{inflation}}}}$. However, it is one order of magnitude more than ${\left( {{\rho _{{B_{{\rm{end}}}}}}} \right)_{{\rm{CMB}}}}$ and much more than ${\left( {{\rho _{{B_{{\rm{end}}}}}}} \right)_{{{\rm{R}}^2} - {\rm{inflation}}}}$. Also, ${\left. {\frac{{d{\rho _B}}}{{d\ln k}}} \right|_{{\rm{end}}}} \gg {\left( {{\rho _{{\rm{reh}}}}} \right)_{{{\rm{R}}^2} - {\rm{inflation}}}},{\left( {{\rho _{{\rm{reh}}}}} \right)_{{\rm{CMB}}}}$. The last result shows that the inflationary magnetogensis may play significant role during the reheating era.  

By taking the magnitude of the magnetic spectra as equivalent to ${\rho _{{B_{{\rm{end}}}}}}$, then the upper bound of magnetic field at the end of inflation, ${B_{{\rm{end}}}} < 3.76 \times {10^{47}}{\rm{G}}$. Therefore, if we use the upper bound of redshift derived from ${\left( {{\rho _{{\rm{end}}}}} \right)_{{\rm{CMB}}}}$, $1 + {z_{{\rm{end}}}} < 6.828 \times {10^{27}}$, the upper limit of the present PMF is, ${B_0} < 8.058 \times {10^{ - 9}}{\rm{G}}$. It is in the same order of magnitude of the upper bound of PMF obtained by Planck, 2015. However, if we use the upper bound of redshift derived from ${\left( {{\rho _{{\rm{end}}}}} \right)_{{{\rm{R}}^2} - {\rm{inflation}}}}$, $1 + {z_{{\rm{end}}}} < 5.414 \times {10^{25}}$, the upper limit of the present PMF will be ${B_0} < 1.282 \times {10^{ - 4}}{\rm{G}}$. That is even higher than the galactic or interplanetary magnetic field which is of the order of ${10^{ - 6}}{\rm{G}}$. Therefore, the second limit is too weak.

\section{Summary}
\label{sec:summary}

In this paper we have shown that the scale invariant PMF can be generated by the simple model ${f^2}FF$, Eq.(\ref{eq:1}), in ${R^2}$-inflationary model, which is mostly favored by the latest result of Planck, 2015 \cite{37}. Unlike, in non-standard inflationary models like NI \cite{39}, LFI \cite{40}, and some standard models of inflation \cite{24} at which ${f^2}FF$ model suffers from the backreaction problem, we can avoid this problem in ${R^2}$-inflation. It is easily to avoid this problem as long as, the rate of inflationary expansion, $H$, is in the order of or less than the upper bound reported by Planck ($ \le 3.6 \times {10^{ - 5}}{M_{{\rm{Pl}}}}$) \cite{37}. In principle, the electric spectrum starts exceeds the magnetic spectrum at relatively high, $H( > 0.2{\rm{ }}{M_{{\rm{Pl}}}})$. The corresponding e-folds number is, $N \sim 68$. Thus, we can consider these two values as upper bounds to the model. We do this investigation for both simple exponential (de Sitter) and power law expansion. At sufficiently high e-folds number, $N$, there is no significant differences in their results.

The second main result is constraining the reheating parameters from the upper limits of PMF reported by Planck, 2015 \cite{9}. In the context of ${R^2}$-inflation, we calculate the lower limits of the reheating parameter, ${R_{{\rm{rad}}}} > 6.888$. Also, we find the other reheating parameters based on the upper limit of energy density at the end of inflation calculated from CBM data, ${\left( {{\rho _{{\rm{end}}}}} \right)_{{\rm{CMB}}}} < 2.789 \times {10^{ - 10}}M_{{\rm{Pl}}}^4$,  and from the inflationary model, ${\left( {{\rho _{{\rm{end}}}}} \right)_{{{\rm{R}}^2} - {\rm{inflation}}}} \simeq {\rm{1}}{\rm{.1}} \times {\rm{1}}{{\rm{0}}^{ - 18}}M_{{\rm{Pl}}}^4$. As a result, we find that the magnetic field energy density at the end of inflation as ${\left( {{\rho _{{B_{{\rm{end}}}}}}} \right)_{{\rm{CMB}}}} < 1.184 \times {10^{ - 20}}M_{{\rm{Pl}}}^4$ and ${\left( {{\rho _{{B_{{\rm{end}}}}}}} \right)_{{{\rm{R}}^2} - {\rm{inflation}}}} < 4.6788 \times {10^{ - 29}}M_{{\rm{Pl}}}^4$. Similarly, we find the upper limit at the end of reheating, ${\left( {{\rho _{{\rm{reh}}}}} \right)_{{\rm{CMB}}}} < 8.480 \times {10^{ - 22}}M_{{\rm{Pl}}}^4$ and ${\left( {{\rho _{{\rm{reh}}}}} \right)_{{{\rm{R}}^2} - {\rm{inflation}}}} < 3.344 \times {10^{ - 30}}M_{{\rm{Pl}}}^4$. All of foregoing results are more than the lower limit derived from WMAP7 for both large and small field inflation \cite{56}. These results show the significance of PMF during reheating era.
   
On the other hand, we constrain the reheating parameters by using the Planck inflationary constraints, 2015 \cite{37} in the context of ${R^2}$-inflation. The upper limit of reheating temperature and energy density for all possible values of ,${w _{{\rm{reh}}}}$ are respectively constrained as, ${T_{{\rm{reh}}}} < 4.32 \times {10^{13}}{\rm{GeV}}$ and ${\rho _{{\rm{reh}}}} < 3.259 \times {10^{ - 18}}M_{{\rm{Pl}}}^4$ at ${n_{\rm{s}}} \approx 0.9674$. This value of spectral index is well consistent with Planck, 2015 results. Adopting ${T_{{\rm{reh}}}}$ for all ${w _{{\rm{reh}}}}$ models of reheating, enables us to constrain the ${N_{{\rm{reh}}}}$ on the range $1 < {N_{{\rm{reh}}}} < 8.3$, for all possible values of ,${w _{{\rm{reh}}}}$. 

At the end, we constrain ${B_0}$, from the scale invariant PMF, generated by ${f^2}FF$ in ${R^2}$-inflation in section. \ref{sec:EM in R2}. From the PMF spectra, Fig.\ref{fig:9}-\ref{eq:10}, we find that the upper limit of magnetic field in the end of inflation is ${B_{{\rm{end}}}} < 3.76 \times {10^{47}}{\rm{G}}$. Therefore, if we use the upper bound of redshift derived from ${\left( {{\rho _{{\rm{end}}}}} \right)_{{\rm{CMB}}}}$, then ${B_0} < 8.058 \times {10^{ - 9}}{\rm{G}}$. It is in the same order of PMF obtained by Planck, 2015. However, if we use the upper bound of redshift derived from ${\left( {{\rho _{{\rm{end}}}}} \right)_{{{\rm{R}}^2} - {\rm{inflation}}}}$, then ${B_0} < 1.282 \times {10^{ - 4}}{\rm{G}}$. That is even higher than the interplanetary or galactic magnetic field which is of the order of ${10^{ - 6}}{\rm{G}}$. Therefore, the second limit is too weak.   

In order to achieve the scale invariant PMF by this model, the free index of the coupling function has a relatively high values, $\alpha  \approx \{  - 7.44,\;{\rm{ }}7.44\} $. However, at $\alpha  = 2$, which is the typical value, we cannot generate scale invariant magnetic field although we can avoid the backreaction problem. The main problem with this model is the value, $\left| \alpha  \right| \approx 7.44$, which is out of the dynamo constraints imposed by CMB, BBN, and Faraday rotation, RM, see Eq.(94) of \cite{24}, see Fig.3 of the same reference. In fact those limits are derived mainly for exponential and large field inflation models. Therefore, this subject needs more investigation on the context of ${R^2}$-inflation.                

If the above problem, the relatively high value of $\alpha $, is not resolved and ${R^2}$-inflation shows more consistency with cosmological observations, the results of this research would go against the simple inflation model ${f^2}FF$, as way of generating PMF. These difficulties add constraints to this model, in addition to those found by other researches, such as new stringent upper limits on the PMF, derived from analyzing the expected imprint of PMF on the CMB power spectra \cite{66}, bi-spectra \cite{67}, tri-spectra \cite{68}, anisotropies and B-modes \cite{69}, and the curvature perturbation and scale of inflation \cite{69}-\cite{70}. However, if there is a way to justify the relatively high $\alpha $ and the approximation of quasi-constant $N$, we assume to derive Eq.\ref{eq:38}, then ${f^2}FF$ model may be viable in the context of ${R^2}$-inflation. Nevertheless, the agreement between the result of this model and the upper limit of present PMF found by Planck, 2015 cast some credits to this model. 

One way to extend this research, is to consider calculating 3-point statistics such as $ \left< BBB \right> $, $\left< BB \phi \right>$ or $\left< BB \zeta \right>$ \cite{71}-\cite{73}. That may enable us to investigate the footprint of the PMF on CMB. It might be addressed in a future research.

\acknowledgments

I would like to thank Rafael Lopez-Mobilia, Bharat Ratra and Robert R. Caldwell for useful discussions.


\begin{thebibliography}{99}


\bibitem{1}
A. H. Guth, \emph{The Infationary Universe: A Possible Solution to the Horizon and Flatness
Problems}, \emph{Phys. Rev.} {\bf D23} (1981) 347.

\bibitem{2}
A. Neronov, and I. Vovk, \emph{Evidence for strong extragalactic magnetic fields from Fermi observations of TeV blazars}, \emph{Science} {\bf 328} (2010) 73.

\bibitem{3}
T. Fujita, and S. Mukohyama, \emph{Universal upper limit on inflation energy scale from cosmic magnetic field} [\href{http://arxiv.org/abs/1205.5031v3}{{\tt arXiv:1205.5031v3}}].

\bibitem{4}
F. Tavecchio et al, \emph{The intergalactic magnetic field constrained by Fermi/LAT observations of the TeV blazar 1ES 0229+200} [\href{http://arxiv.org/abs/1004.1329v2}{{\tt arXiv:1004.1329v2}}].

\bibitem{5}
K. Ichiki, K. Takahashi,  and N. Sugiyama, \emph{Constraint on the primordial vector mode and its magnetic field generation from seven-year Wilkinson Microwave Anisotropy Probe Observations} [\href{http://arxiv.org/abs/1112.4705v1}{{\tt arXiv:1112.4705v1}}].

\bibitem{6}
P. A. R. Ade et al, \emph{Planck intermediate results. XXXIII. Signature of the magnetic field geometry of interstellar filaments in dust polarization maps} [\href{http://arxiv.org/abs/1411.2271v1}{{\tt arXiv:1411.2271v1}}].

\bibitem{7}
W. Esseya, S. Andob, and A. Kusenko, \emph{Determination of intergalactic magnetic fields from gamma ray data}, \emph{Astroparticle Physics} {\bf 35} (2011) 135 [\href{http://arxiv.org/abs/1012.5313}{{\tt arXiv:1012.5313}}].

\bibitem{8}
T. Kahniashvili et al, \emph{Primordial magnetic field limits from cosmological data}, \emph{Phys. Rev.} {\bf D82} (2010) 083005 [\href{http://arxiv.org/abs/1012.5313}{{\tt arXiv:1012.5313}}].

\bibitem{9}
P. A. R. Ade et al, \emph{Planck 2015 results. XIX. Constraints on primordial magnetic fields} [\href{http://arxiv.org/abs/1502.01594v1}{{\tt arXiv:1502.01594v1}}].

\bibitem{10}
K. Subramanian, and J. Barrow, \emph{Magnetohydrodynamics in the Early Universe and the Damping of Non-linear Alfven Waves}, \emph{Phys. Rev.} {\bf D58} (1998) 135 [\href{http://arxiv.org/abs/9712083v1}{{\tt arXiv:astro-ph/9712083v1}}].

\bibitem{11}
K. Jedamzik, V. Katalini´c and A. Olinto, \emph{Damping of Cosmic Magnetic Fields}, \emph{Phys. Rev.} {\bf D57} (1998) 3264 [\href{http://arxiv.org/abs/9606080v2}{{\tt  arXiv:astro-ph/9606080v2}}].

\bibitem{12}
D. Moss and D. Sokoloff, \emph{Stringent magnetic field limits from early universe dynamos cosmology with torsion} [\href{http://arxiv.org/abs/1307.0142v1}{{\tt arXiv:1307.0142v1}}].

\bibitem{13}
D. Grasso and H. Rubinstein, \emph{Magnetic Fields in the Early Universe}, \emph{Phys. Rept.} {\bf 348} (2001) 163 [\href{http://arxiv.org/abs/0009061v2}{{\tt arXiv:astro-ph/0009061v2}}].

\bibitem{14}
L. Widrow, \emph{Origin of Galactic and Extragalactic Magnetic Fields}, \emph{Rev. Mod. Phys.} {\bf 74} (2002) 775 [\href{http://arxiv.org/abs/0207240v1}{{\tt arXiv:astro-ph/0207240v1}}].

\bibitem{15}
J. Barrow, R. Maartens, and C. Tsagas, \emph{Cosmology with inhomogeneous magnetic fields}, \emph{Phys. Rept.} {\bf 449} (2007) 131 [\href{http://arxiv.org/abs/0611537v4}{{\tt arXiv:astro-ph/0611537v4}}].

\bibitem{16}
A. Kandus, K. Kunze, and C. Tsagas, \emph{Primordial magnetogenesis}, \emph{Phys. Rept.} {\bf 348} (2001) 163 [\href{http://arxiv.org/abs/1007.3891v2}{{\tt arXiv:1007.3891v2}}].

\bibitem{17}
D. Ryu et al, \emph{Magnetic Fields in the Large-Scale Structure of the Universe}, \emph{Space. Sci. Rev.} {\bf 166} (2012) 1 [\href{http://arxiv.org/abs/1109.4055v1}{{\tt  arXiv:1109.4055v1}}].

\bibitem{18}
L. Widrow et al, \emph{The First Magnetic Fields}, \emph{Space. Sci. Rev.} {\bf 166} (2012) 37 [\href{http://arxiv.org/abs/1109.4052v1}{{\tt arXiv:1109.4052v1}}].

\bibitem{19}
D.Yamazaki et al, \emph{The Search for a Primordial Magnetic Field}, \emph{Phys. Rept.} {\bf 517} (2012) 141 [\href{http://arxiv.org/abs/0009061v2}{{\tt arXiv:astro-ph/0009061v2}}].

\bibitem{20}
R. Durrer and A. Neronov, \emph{Cosmological Magnetic Fields: Their Generation, Evolution and Observation} [\href{http://arxiv.org/abs/1303.7121v2}{{\tt arXiv:1303.7121v2}}].

\bibitem{21}
M. S. Turner and L. M. Widrow, \emph{Inflation-produced, large-scale magnetic fields}, \emph{Phys. Rev.} {\bf D37} (1988) 2743.

\bibitem{22}
B. Ratra, \emph{Inflation Generated Cosmological Magnetic Field},(1991), GRP-287/CALT-68-1751.

\bibitem{23}
B. Ratra, \emph{Cosmological seed magnetic field from inflation}, \emph{Astrophys. J.} {\bf 391} (1992) L1.

\bibitem{24}
J. Martin and J. Yokoyama, \emph{Generation of large scale magnetic fields in single-field inflation}, \emph{JCAP} {\bf 0801} (2008) 025 [\href{http://arxiv.org/abs/0711.4307v1}{{\tt arXiv:0711.4307v1}}].

\bibitem{25}
K. Subramanian, \emph{Magnetic fields in the early universe}, \emph{Astron.Nachr.} {\bf 331} (2012) 110 [\href{http://arxiv.org/abs/0911.4771v2}{{\tt arXiv:0911.4771v2}}].

\bibitem{26}
B. Himmetoglu, \emph{Vector Fields During Cosmic Inflation: Stability Analysis and Phenomenological Signatures}, University of Minnesota Digital Conservancy (2010) [\href{http://purl.umn.edu/93920}{{\tt http://purl.umn.edu/93920}}].

\bibitem{27}
V. Demozzi, V. Mukhanov, and H. Rubinstein, \emph{Magnetic fields from inflation}, \emph{JCAP} {\bf 0908} (2009) 025 [\href{http://arxiv.org/abs/0907.1030v1}{{\tt arXiv:0907.1030v1}}].

\bibitem{28}
R. J. Z. Ferreira, R. K. Jain, and M. S. Sloth, \emph{Inflationary Magnetogenesis without the Strong Coupling Problem}, \emph{JCAP} {\bf 10} (2013) 004 [\href{http://arxiv.org/abs/1305.7151v3}{{\tt arXiv:1305.7151v3}}].

\bibitem{29}
A. Linde, \emph{Chaotic inflating universe}, \emph{Phys. Lett.} {\bf B129} (1983) 177.

\bibitem{30}
J. Martin, C. Ringeval, and V. Vennin, \emph{Encyclopdia Infationaris}, \emph{Phys. Dark Univ.}{\bf 5} (2014) 75 [\href{http://arxiv.org/abs/1303.3787v3}{{\tt arXiv:1303.3787v3}}].

\bibitem{31}
F. Lucchin and S. Matarrese, \emph{Magnetic fields in the early universe}, \emph{Phys. Rev.} {\bf D32} (1985) 1316.

\bibitem{32}
E. Komatsu et al, \emph{Seven-Year Wilkinson Microwave Anisotropy Probe (WMAP) Observations: Cosmological Interpretation}, \emph{Astrophys. J. Suppl.} {\bf 192} (2011) 18 [\href{http://arxiv.org/abs/1001.4538v3}{{\tt arXiv:1001.4538v3}}].

\bibitem{33}
G. Hinshaw et al, \emph{Nine-Year Wilkinson Microwave Anisotropy Probe (WMAP) Observations: Cosmological Parameter Results}, \emph{Astrophys. J. Suppl} {\bf 208} (2013) 19 [\href{http://arxiv.org/abs/1212.5226v3}{{\tt arXiv:1212.5226v3}}].

\bibitem{34}
P. A. R. Ade et al, \emph{Planck 2013 results. XXII. Constraints on inflation}, \emph{Astron. Astrophys} {\bf 571} (2014) A22 [\href{http://arxiv.org/abs/1303.5082v2}{{\tt arXiv:1303.5082v2}}].

\bibitem{35}
A. Linde, \emph{Inflationary Cosmology after Planck 2013} [\href{http://arxiv.org/abs/1402.0526v2}{{\tt arXiv:1402.0526v2}}].

\bibitem{36}
P. A. R. Ade et al, \emph{A Joint Analysis of BICEP2/Keck Array and Planck Data 2015}, \emph{Phys. Rev. Lett.} {\bf 114} (2015) 101301 [\href{http://arxiv.org/abs/1502.00612v1}{{\tt  arXiv:1502.00612v1}}].

\bibitem{37}
P. A. R. Ade et al, \emph{Planck 2015. XX. Constraints on inflation} [\href{http://arxiv.org/abs/1502.02114v1}{{\tt arXiv:1502.02114v1}}].

\bibitem{38}
P. A. R. Ade et al, \emph{Detection Of B-mode Polarization at Degree Angular Scales}, \emph{Phys. Rev. Lett.} {\bf 112} (2014) 241101 [\href{http://arxiv.org/abs/1403.3985}{{\tt arXiv:1403.3985}}].

\bibitem{39}
A. AlMuhammad, R. Lopez-Mobilia, \emph{The Early Universe ${f^2}FF$ Model of Primordial Magnetic Field at Natural Inflation}. (submitted to publication)

\bibitem{40}
A. AlMuhammad, R. Lopez-Mobilia, \emph{The Early Universe ${f^2}FF$ Model of Primordial Magnetic Field at Large Field Inflation}. (submitted to publication)

\bibitem{41}
A. A. Starobinsky, \emph{A New Type of Isotropic Cosmological Models Without Singularity}, \emph{Phys. Lett.} {\bf B91} (1980) 99.

\bibitem{42}
F. Bezrukov and M. Shaposhnikov, \emph{Detection Of B-mode Polarization at Degree Angular Scales}, \emph{Phys. Lett.} {\bf B659} (2008) 703 [\href{http://arxiv.org/abs/0710.3755v2}{{\tt arXiv:0710.3755v2}}].

\bibitem{43}
A. R. Liddle and D. H. Lyth, \emph{Cosmological Inflation and Large-Scale Structure}, Cambridge University Press (2000).

\bibitem{44}
A. R. Liddle, P. Parsons and J. D. Barrow, \emph{Formalising the Slow-Roll Approximation in Inflation}, \emph{Phys. Rev.} {\bf D50} (1994) 7222 [\href{http://arxiv.org/abs/9408015v1}{{\tt arXiv:astro-ph/9408015v1}}].

\bibitem{45}
J. Martin, and D. Schwarz, \emph{The precision of slow-roll predictions for the CMBR anisotropies}, \emph{Phys. Rev.} {\bf D62} (2000) 103520 [\href{http://arxiv.org/abs/9911225v2}{{\tt arXiv:astro-ph/9911225v2}}].

\bibitem{46}
V. Mukhanov, \emph{Physical Foundation of Cosmology}, Cambridge University Press (2005).

\bibitem{47}
M. S. Turner, \emph{Coherent scalar field oscillations in an expanding universe}, \emph{Phys. Rev.} {\bf D28} (1983) 1243.

\bibitem{48}
L. Kofman, A. D. Linde and A. A. Starobinsky, \emph{Towards the Theory of Reheating After Inflation}, \emph{Phys. Rev.} {\bf D56} (1997) 3258 [\href{http://arxiv.org/abs/9704452}{{\tt arXiv:hep-ph/9704452}}].

\bibitem{49}
B. A. Bassett, S. Tsujikawa and D. Wands, \emph{Inflation Dynamics and Reheating}, \emph{Rev. Mod. Phys.} {\bf 78} (2006) 537 [\href{http://arxiv.org/abs/0507632}{{\tt arXiv:astro-ph/0507632}}].

\bibitem{50}
F. Finelli and R. H. Brandenberger, \emph{Parametric Amplification of Gravitational Fluctuations During Reheating}, \emph{Phys. Rev. Lett.} {\bf  82} (1999) 1362 [\href{http://arxiv.org/abs/9809490v2}{{\tt arXiv:hep-ph/9809490v2}}].

\bibitem{51}
B. A. Bassett, D. I. Kaiser and R. Maartens, \emph{General Relativistic effects in preheating}, \emph{Phys. Lett.} {\bf B455} (1999) 84 [\href{http://arxiv.org/abs/9808404}{{\tt arXiv:hep-ph/9808404}}].

\bibitem{52}
F. Finelli and R. H. Brandenberger, \emph{Parametric Amplification of Metric Fluctuations During Reheating in Two Field Models}, \emph{Phys. Rev.} {\bf D62} (2000) 083502 [\href{http://arxiv.org/abs/0003172}{{\tt arXiv:hep-ph/0003172}}].

\bibitem{53}
K. Jedamzik, M. Lemoine and J. Martin, \emph{Collapse of Small-Scale Density Perturbations during Preheating in Single Field Inflation}, \emph{JCAP} {\bf 1009} (2010) 034 [\href{http://arxiv.org/abs/1002.3039v1}{{\tt  arXiv:1002.3039v1}}].

\bibitem{54}
K. Jedamzik, M. Lemoine and J. Martin, \emph{Generation of gravitational waves during early structure formation between cosmic inflation and reheating}, \emph{JCAP} {\bf 1004} (2010) 021 [\href{http://arxiv.org/abs/1002.3278}{{\tt arXiv:1002.3278}}].

\bibitem{55}
R. Easther, R. Flauger and J. B. Gilmore, \emph{Delayed Reheating and the Breakdown of Coherent Oscillations} [\href{http://arxiv.org/abs/1003.3011}{{\tt arXiv:1003.3011}}].

\bibitem{56}
J. Martin and C. Ringeval, \emph{First CMB Constraints on the Inflationary Reheating Temperature}, \emph{Phys. Rev.} {\bf D82} (2010) 023511 [\href{http://arxiv.org/abs/1004.5525v2}{{\tt arXiv:1004.5525v2}}].

\bibitem{57}
M. Bastero-Gil, V. Di Clemente and S. F. King, \emph{Large Scale Structure from the Higgs fields of the Supersymmetric Standard Model}, \emph{Phys. Rev.} {\bf D67} (2003) 103516 [\href{http://arxiv.org/abs/0211011}{{\tt arXiv:hep-ph/0211011}}].

\bibitem{58}
Mazumdar and J. Rocher, \emph{Particle physics models of inflation and curvaton scenarios}, \emph{Phys. Rept.} {\bf 497} (2011) 85 [\href{http://arxiv.org/abs/1001.0993}{{\tt arXiv:1001.0993}}].

\bibitem{59}
V. Demozzi and C. Ringeval, \emph{Reheating constraints in inflationary magnetogenesis}, \emph{JCAP} {\bf 1205} (2012) 009 [\href{http://arxiv.org/abs/1202.3022v2}{{\tt arXiv:1202.3022v2}}].

\bibitem{60}
A. R. Liddle and S. M. Leach, \emph{How long before the end of inflation were observable perturbations produced?}, \emph{Phys. Rev.} {\bf D68} (2003) 103503 [\href{http://arxiv.org/abs/0305263}{{\tt arXiv:astro-ph/0305263}}].

\bibitem{61}
J. Martin and C. Ringeval, \emph{Inflation after WMAP3: Confronting the Slow-Roll and Exact Power Spectra with CMB Data}, \emph{JCAP} {\bf 0608} (2006) 009 [\href{http://arxiv.org/abs/0605367v2}{{\tt arXiv:astro-ph/0605367v2}}].

\bibitem{62}
J. L. Cook et al, \emph{Reheating predictions in single field inflation}, \emph{JCAP} {\bf 1009} (2010) 034 [\href{http://arxiv.org/abs/1502.04673v1}{{\tt arXiv:1502.04673v1}}].

\bibitem{63}
J. B. Munoz and M. Kamionkowski, \emph{Equation-of-State Parameter for Reheating}, \emph{Phys. Rev.} {\bf D 91} (2015) 043521 [\href{http://arxiv.org/abs/1412.0656v2}{{\tt arXiv:1412.0656v2}}].

\bibitem{64}
J. Martin and C. Ringeval, \emph{Inflation after WMAP3: Confronting the Slow-Roll and Exact Power Spectra with CMB Data}, \emph{JCAP} {\bf 08} (2006)009 [\href{http://arxiv.org/abs/0605367}{{\tt astro-ph/0605367}}].

\bibitem{65}
C. Ringeval, T. Suyama and J. Yokoyama, \emph{Magneto-reheating constraints from curvature perturbations}, \emph{Phys. Rept.} {\bf 497} (2011) 85 [\href{http://arxiv.org/abs/1302.6013v1}{{\tt arXiv:1302.6013v1}}].

\bibitem{66}
P. A. R. Ade et al, \emph{Planck 2013 results. XXII. Constraints on inflation}, \emph{Astron. Astrophys.} {\bf 571} (2014) A22 [\href{http://arxiv.org/abs/1303.5082v2}{{\tt arXiv:1303.5082v2}}].

\bibitem{67}
P. Trivedi, K. Subramanian and T. R. Seshadri, \emph{Primordial Magnetic Field Limits from Cosmic Microwave Background Bispectrum of Magnetic Passive Scalar Modes}, \emph{Phys. Rev.} {\bf D82} (2010) 123006 [\href{http://arxiv.org/abs/1009.2724v1}{{\tt arXiv:1009.2724v1}}].

\bibitem{68}
P. Trivedi, K. Subramanian and T. R. Seshadri, \emph{Primordial Magnetic Field Limits from CMB Trispectrum - Scalar Modes and Planck Constraints}, \emph{Phys. Rev} {\bf D89} (2014) 043523 [\href{http://arxiv.org/abs/0907.1030v1}{{\tt arXiv:1312.5308v1}}].

\bibitem{69}
R. Z. Ferreira, R. K. Jain and M. S. Sloth, \emph{Inflationary Magnetogenesis without the Strong Coupling Problem II: Constraints from CMB anisotropies and B-modes}, \emph{JCAP} {\bf 1406} (2014) 053 [\href{http://arxiv.org/abs/1403.5516v2}{{\tt arXiv:1403.5516v2}}].

\bibitem{70}
T. Fujitaa and Sh. Yokoyama, \emph{Critical constraint on inflationary magnetogenesis} [\href{http://arxiv.org/abs/1402.0596}{{\tt arXiv:1402.0596}}].

\bibitem{71}
R. R. Caldwell, L. Motta and M. Kamionkowski, \emph{Correlation of inflation-produced magnetic fields with scalar fluctuations}, \emph{Phys. Rev} {\bf D84} (2011) 123525 [\href{http://arxiv.org/abs/1109.4415}{{\tt arXiv:1109.4415}}].

\bibitem{72}
L. Motta and R. R. Caldwell, \emph{Non-Gaussian features of primordial magnetic fields in power-law inflation}, \emph{Phys. Rev} {\bf D85} (2012) 103532 [\href{http://arxiv.org/abs/1203.1033}{{\tt arXiv:1203.1033}}].

\bibitem{73}
T Fujita et al, \emph{Consistent generation of magnetic fields in axion inflation models} [\href{http://arxiv.org/abs/1503.05802}{{\tt arXiv:1503.05802}}].




% Please avoid comments such as "For a review'', "For some examples",
% "and references therein" or move them in the text. In general,
% please leave only references in the bibliography and move all
% accessory text in footnotes.

% Also, please have only one work for each \bibitem.


\end{thebibliography}
\end{document}